\input harvmac
\input epsf
$\,$
\def\CM{{\cal M}}
\font\cmss=cmss10 \font\cmsss=cmss10 at 10truept
\def\IR{\relax{\rm I\kern-.18em R}}
\font\cmss=cmss10 \font\cmsss=cmss10 at 10truept
\def\IZ{\relax\ifmmode\mathchoice
{\hbox{\cmss Z\kern-.4em Z}}{\hbox{\cmss Z\kern-.4em Z}}
{\lower.9pt\hbox{\cmsss Z\kern-.36em Z}}
{\lower1.2pt\hbox{\cmsss Z\kern-.36em Z}}\else{\cmss Z\kern-.4em Z}\fi}
\overfullrule=0pt
\Title{\vbox{\baselineskip12pt\hbox{DFTUZ  93.10\phantom{lll}}\hbox{}}}
{\vbox{\centerline{Universality and Ultraviolet
Regularizations}\medskip \centerline{of Chern-Simons Theory}}}

\centerline{\bf M. Asorey, F. Falceto,
J. L. L\'opez and G. Luz\'on}
\bigskip\centerline{{ Departamento de
F\'{\i}sica Te\'orica. Facultad de Ciencias}}
\centerline{Universidad de Zaragoza.  50009 Zaragoza. Spain}

\def\cs{Chern-Simons}
\def\cst{Chern-Simons theory}

\baselineskip=16pt plus 2pt minus 1pt
\bigskip
\bigskip
\medskip
\centerline {\bf Abstract}
  	The universality of
radiative corrections to the gauge coupling constant $k$ of Chern-Simons
theory is studied in a very general regularization scheme. We show that
the effective coupling constant $k$ induced by radiative corrections
depends crucially on the balance between the ultraviolet behavior
of
scalar and pseudoscalar terms in the regularized action. There
are three different regimes. When the ultraviolet leading term
is scalar the coupling $k$ is shifted to $k+h^{\vee}$.
However, if the leading term is pseudoscalar the shift is
 $k+s h^{\vee}$ with $s=0$ or $s=2$ depending on the
sign of such a term. In
the borderline case when
the scalar  and pseudoscalar terms have the same ultraviolet
behavior the shift of $k$ becomes arbitrary (even non-integer)
and depends on the parameters of the regularization. We also
show that the coefficient of the induced gravitational
Chern-Simons term is different for  the three regimes and has the
same universality properties  than the effective coupling
constant $k$. The results open the possibility of a
connection  with non-rational two-dimensional conformal
theories in  the borderline regime.

 \overfullrule=0pt  \hyphenation{systems}
\bigskip\vfill \noindent\baselineskip=12pt plus 2pt minus 1pt
\Date{ }

\parindent=20pt
\baselineskip=16pt plus 2pt minus 1pt
\newsec{Introduction}
The invariance of Chern-Simons theory under  gauge
transformations imposes a quantization condition   on its coupling
constant $k\in \IZ$  when the gauge group is compact.  This
constraint arises in the
covariant formalism
as a consistency condition for the definition of the
euclidean functional integral
due to the special transformation properties
of the Chern-Simons action under
large gauge transformations \ref\jdt{
 S. Deser, R. Jackiw, S. Templeton, Phys. Rev. Lett. { 48}
(1982)    975;   Ann. Phys. (N.Y.) { 140} (1982) 372}.
In the canonical formalism it appears as a necessary condition for the
integration of Gauss law constraint on the physical states \ref\runo
{R. Jackiw,
In { \it Gauge Theories of the Eighties}, Eds. E.
Ratio, J. Lindfords, Lecture Notes in Physics, vol. 181, Springer
(1983) \semi
M. Asorey, P.K. Mitter, Phys. Lett. B153 (1985) 147}.
Both interpretations are based on
non-infinitesimal symmetries and therefore the quantization
condition can not be derived from perturbative arguments. However,
unexpectedly the perturbative contributions of quantum
fluctuations do not seem
to change the integer nature of the  Chern-Simons coupling
constant in the renormalization  schemes considered so far.
Regularization methods  which do preserve the pseudoscalar character
of Chern-Simons interaction do not yield any renormalization of the
coupling constant \ref\ita{
E. Guadagnini, M. Martellini, M. Mintchev, Phys. Lett. { B 227}
(1989) 111}\ref\rlc{M. Asorey, F. Falceto,
J.L. L\'opez and G. Luz\'{o}n,
 Phys. Rev. {\bf B 49} (1993) 5377}
and the regularizations which introduce  scalar
Yang-Mills like selfinteractions generate by one loop corrections
an integer shift of the coupling constant of the form
$k\rightarrow k+h^{\vee}$\ref\pisrao{R. D. Pisarski, S. Rao,
 Phys. Rev.  D 32 (1985) 2081}\nref\sem{W.
Chen,  G. W. Semenoff, Y.-S. Wu, Mod. Phys. Lett. A5  (1990)
1833}\nref\LAG{L. Alvarez-Gaum\'e, J.M.F. Labastida and A.V.
Ramallo, Nucl. Phys. {
B334} (1990) 103}\nref\rafa{ M. Asorey, F. Falceto,
Phys. Lett. B 241 (1990) 31}--\ref\carmelo{
C.P.  Martin, Phys.  Lett. { B 241} (1990) 513
\semi
M. A. Shifman, Nucl. Phys. B352 (1991) 87
}. The
behavior of  the effective coupling constant $k$
with these two types of regularizations seems to
indicate the existence of an universal property of $k$ perhaps
related to the topological nature of the Chern-Simons
theory\foot{
A similar phenomenon occurs for non-compact gauge groups
\ref\nat{D.
Bar--Natan, E. Witten, Commun. Math. Phys. { 141}(1991) 423}}.
However from a pure quantum field theory point of view this
behavior is unusual because in absence of perturbative symmetry
constraints \rafa\ref\sus{F. Delduc, C. Lucchesi, O. Piguet and S.P. Sorella,
Nucl. Phys.  B346 (1990) 313}   there must always exist a
regularization scheme where the effective values
of coupling constants are arbitrary. In
a recent paper \rlc\
we have found  by a suitable introduction of
pseudoscalar interactions with Pauli-Villars ghosts that
the effective
value of $k$ is not constrained to be only  $k$ or $k+h^{\vee}$ but
$k+nh^{\vee}$, where  $n$ is an arbitrary integer  which
depends on the number of ghosts fields  interacting with
gauge fields by
means of pseudoscalar couplings
\foot{This result is reminiscent of that obtained by
 Coste-L\" usher for the coefficient of the Chern-Simons  term
in the effective action generated by Dirac fermions in $2+1$
space-time dimensions}.
However, even within this general scheme the integer character of
the shift is preserved and this property
requires a physical explanation.

In this paper we find a physical interpretation of this phenomenon
 by considering the quantization of \cst\  in a more  general
regularization scheme. The regularization involves pseudoscalar and
scalar terms with arbitrary number of covariant derivatives
in such a way that gauge invariance is preserved. We observe
an interplay between the ultraviolet behaviors of scalar and
pseudoscalar terms, which generates  three different
regimes for the radiative
corrections generated by loops of gluons. In the first regime
the leading ultraviolet selfinteractions of gluons are scalar. The
effective \cs\
coupling constant gets  shifted in this case by  the dual Coxeter
number $h^{\vee}$ of $G$, i.e. $k\rightarrow k+h^{\vee}$, due to
one loop gluonic radiative corrections. The second regime is
characterized by an ultraviolet behavior dominated by pseudoscalar
selfinteractions and the absence of radiative
corrections to $k$ if the sign of those leading terms match that of
the Chern-Simons term. If the signs of those terms are opposite,
then, the radiative corrections shift $k$ to  $k+ 2
h^{\vee}$. In the borderline regime scalar and pseudoscalar
terms have the same ultraviolet behavior. In such a case the
quantum
corrections to $k$ can take any real value
and do depend  on the relative coefficients of the
leading terms of scalar and pseudoscalar interactions.

The  universality properties of the above three regimes
are not only present in the renormalization of the coupling constant
$k$.
They also appear in other observables of the quantum
theory, e.g.
in the form of the induced gravitational action. The one-loop
contribution to coefficient $\kappa$ of the induced  gravitational
Chern-Simons
term  is $\dim G/24$ in the regime dominated by  scalar
interactions,
vanishes or is equal to $\dim G/12$ in the axial regime
and depends on  the relative coefficients of
the leading terms of scalar and axial interactions like
the shift of $k$ in the borderline regime. This features indicate
that the three regimes correspond in fact to different physical
behaviors, and point out   the physical relevance of the
balance of the ultraviolet behaviors of the
axial and scalar regulating terms.

 The structure of the paper is as follows. In section 2 we
introduce the most general regularization within the framework of
geometric regularizations which will be used throughout the
paper. The consistency of the regularization is analyzed in
section $3$. This  imposes some restrictions on the nature of
ghosts interactions. In particular,
finiteness and gauge
invariance can only be achieved, for regularizations with
leading
axial terms in the ultraviolet regime,  by means of pseudo-differential
non-local
operators in the interaction of ghosts with gluons.
In section $4$ we carry out the calculation of one-loop corrections
to the effective action. We
observe the  three regimes  already mentioned  leading to different
quantum corrections. In most  cases there is a non-vanishing
renormalization of the gauge field, first pointed out in
\rafa.
In section $5$ we analyse the nature of the induced gravitational
action  and in particular we evaluate the coefficient of the
gravitational \cs\ term which is supposed to be proportional to the
central charge
of the associated Wess-Zumino-Witten model. Finally in section $6$
we
summarize the conclusions and implications of the present study.
Some
technical aspects concerning perturbative calculations are
summarized in the two appendices.

\newsec{Ultraviolet Regularization}

Because of the pseudoscalar character of the Chern-Simons
action, standard  perturbative regularization methods can not be
applied. This fact, has recently stimulated  the  interest
on the application of different perturbative regularization
prescriptions to Chern-Simons theories
\ita--\ref\dimn{G. Giavarini, C.P. Martin,  F. Ruiz, Nucl.Phys.
{ B 381}(1992) 222}.
We will consider an ultraviolet regularization based on the
geometric regularization scheme introduced by two of us for gauge
theories in ref. \ref\gr{M. Asorey, F. Falceto, Phys. Lett. B 206
(1988) 485; Nucl. Phys. B 32 (1989) 427}. The essential features of
this type of regularization are based on the observation that the
relevant space for covariant quantization is the
space  of gauge fields modulo  gauge transformations, i.e.
the space of gauge orbits
 $\CM$, endowed with a Riemannian volume element \ref\bab{
O. Babelon, C. M. Viallet, Phys. Lett. B85 (1979) 246}.
Since $\CM$ is a curved $\infty$-dimensional (Riemannian) manifold
the regularization of a functional integral defined over $\CM$ does
not simply requires a regularization of the action, as in ordinary
field theories with flat configuration spaces, but also a
non-trivial regularization of the functional (Riemannian) volume
element \ref\AM{
M. Asorey, P.K. Mitter, Commun. Math. Phys. 80 (1981) 43
}. In this way it is possible to obtain
a regularization which overcomes the Gribov problem and is free
of  overlapping divergences usually associated to the
regularization by means of higher covariant derivatives and
Pauli-Villars ghosts, whereas it preserves all topological
properties of continuum approaches and has a non-perturbative
interpretation (see Refs.
\gr\ ).

The geometric regularization method proceeds by three steps.

1.  {\it Regularization of the classical  action}
 by  higher covariant derivatives. There are two classes of
regulating terms which preserve gauge invariance:
 scalar (real) terms of Yang-Mills type \foot{A pure Yang-Mills
term is not sufficient to regulate 2-loops divergences
\LAG.}
, e.g.
\eqn\euno{S^s_\Lambda = {k\over 8\pi \Lambda}
(F(A),(I + {\Delta_A/ \Lambda^{^2}})^{m} F(A))
}
with $\Delta_A = d^{\ast }_A\, d^{\, }_A +
d^{\, }_A\, d^{\ast }_A$ and pseudoscalar
(axial, imaginary) terms, e.g.
\eqn\euno{S^a_\Lambda =-{i k\over 8\pi \Lambda^2}
(\ast F(A),(I + {\Delta_A/ \Lambda^{^2}})^{n}\ast d^{\, }_A
(I + {\Delta_A/ \Lambda^{^2}})^{n} \ast F(A))
}
which were first introduced in ref. \rafa\ for different purposes.
We shall consider the most general
gauge invariant structure of the regularized action
\eqn\euno{S_\Lambda= S^{\rm cs}+ \lambda
S^s_\Lambda + \lambda'  S^a_\Lambda }
involving both types of regulating terms with relative (real)
weights
$\lambda$  $\lambda'$  in addition to the Chern-Simons
action
\eqn\ecs{S^{\rm cs}={i k\over 4\pi }\int_{T^3}\ tr\
(A\wedge d A + {2\over 3} A\wedge
A\wedge A).}
 We assume that $\lambda\geq 0$ to have a damping contribution
to the functional integral and use the
compact notation  introduced in ref. \gr\ for  covariant
gauge differential calculus.

2.  {\it Regularization of the volume   of gauge orbits},
$det^{^{1/2}}   \Delta^{^0}_A=det^{^{1/2}}d^\ast_{A}d^{\ }_{A}$,
by Pauli-Villars method
\eqn\ecinco{\eqalign{
det^{ }_{\Lambda}\Delta^{^0}_{A} =&
  det^{^{-1}}(I +
 \Delta^{^0}_A/ {\Lambda}^{^2})^{2m_0}\
  det^{^{-1}}(I +
 \Delta^{^0}_A/ {\Lambda}^{^2})^{2(m_1-m_0)}\
  det^{^{-1}}(I +
 \Delta^{^0}_A/ {\Lambda}^{^2})^{2(m_2-m_1)}\cr
& det^{^{}}\Delta^{^0}_{A}\left([(I+ \lambda'(I +
  \Delta^{^0}_A/
\Lambda^{^2})^{2n}\Delta^{^0}_{A}/\Lambda^2)]^2+ \lambda^2(I +
  \Delta^{^0}_A/
\Lambda^{^2})^{2m}\Delta^{^0}_{A}/\Lambda^2\right)\cr
}}
and the crucial step

3. {\it Regularization of the volume element
of the  gauge orbit space} $ \CM = \CAG$. In the case
of scalar fields the
configuration space  has a Hilbert space structure $\CH$ and
  a Gaussian measure in $\CH$  is defined
by means of the nuclear structure associated to  a trace class
operator (Minlos' theorem). The
generalization of this structure for $\infty$-dimensional curved
spaces
is also necessary for
the construction of functional measures on Hilbert Lie  groups
\ref\Dal{Ju. Daletskii,  Russ. Math. Surv. 22 (1967) 1\semi
Ju. L. Daletskii, Ya. I. Shnaiderman, Funct. Anal. Appl. 3
(1969) 88} and arbitrary  $\infty$-dimensional Hilbert
manifolds including gauge orbits spaces
 \AM. Geometric regularization is based on  the
implementation of this construction for the covariant formalism
\gr.
This is achieved  by means   of a binuclear Riemannian
structure $(g^{0 }_{\Lambda} , G^{^1}_{\Lambda},
G^{^2}_{\Lambda})$
  of $\CM$, which  consists of a Riemannian metric
$g^{0 }_{\Lambda}$ and two
families of selfadjoint trace class operators
$G^{^1}_{\Lambda}, G^{^2}_{\Lambda}$ acting on the
tangent spaces of $\CM$.
In our case the binuclear
Riemannian  structure can be defined by means of  the  three
Riemannian metrics $g^{i }_{\Lambda}, i=0,1,2 $ of $\CM$  given by
\eqn\ediez{
g^{^i}_{\Lambda} (\tau,\eta) = (P_{A}\widetilde\tau,(I +
{\Delta_A/{\Lambda}^{^2}})^{m_i}P_{A} \widetilde\eta)\qquad
i=0,1,2,
}
for any tangent vectors $\widetilde\tau,\widetilde\eta$ of $T_A\CA$
whose projections on $\CM$ are $\tau$ and $\eta$, respectively.
$P_{A}$ denotes the projection operator
\eqn\eseis {P_{A} = 1-d^{\ }_{A} (d^\ast_{A} d^{\
}_{A})^{-1} d^\ast_{_{A}}}
onto the subspace of  tangent vectors
 $\xi \in T_{A}\CA$ which are transverse to the gauge orbit of
$A$, $d^{\ast}_{A} {\xi}=0$.
The operators
 $G^i_{\Lambda} : T_{[A]}{\CM}\longrightarrow T_{[A]}{\CM }
\  (i=1,2)$ defined by
\eqn\eseis{
 g^{^i}_{\Lambda} (\tau,\eta) = g^{0}_{\Lambda}
(\tau,(G^{^i}_{\Lambda})^{-1}\eta)
}
are selfadjoint and
trace class with respect to $ g^{0 }_{\Lambda}$   for
$m_i\geq m_0+ 2$, and define a binuclear Riemannian  structure in
the orbit space  $\CM$.

In the local coordinates of $\CM$ defined by the
gauge condition
\eqn\esos {d^\ast A_c  = 0}
 in a neighborhood $\CC_{0}$ of the orbit of $A=0$, the
regularized functional integral reads
\eqn\cinco {\int_{\CC_{0}}  \delta  A_c  \ det_c^{^{1/2}}
g^{0 }_\Lambda
(A_c)
 \ det_c^{^{1/2}} (G^{^1}_\Lambda)^{^{-1}}  \ det_c^{^{1/2}}
(G^{^2}_\Lambda)^{^{-1}} G^{^1}_\Lambda\  det_\Lambda^{^{1/2}}
\Delta^{^0}_A\, \ {\rm
e}^{-S_{\Lambda}(A)},}
where the  functional formal measure
\eqn\seis {\delta \mu_{ g^{0}_\Lambda  G^{^1}_\Lambda
G^{^2}_\Lambda} (A_c) = \delta  A_c  \ det_c^{^{1/2}}g^{0 }_\Lambda
(A_c)
 \ det_c^{^{1/2}} (G^{^1}_\Lambda)^{^{-1}}  \ det_c^{^{1/2}}
(G^{^2}_\Lambda)^{^{-1}} G^{^1}_\Lambda.}
can be considered as a regularization of the volume element
associated to the binuclear riemannian structure $(g^{0}_\Lambda ,
G^{^1}_\Lambda, G^{^2}_\Lambda)$.

We remark that the definition of the regularization is based on pure
geometrical terms and makes sense beyond perturbation theory. Therefore
 the
regularized functional integral
\eqn\siete{\int_{\CM}\delta \mu_{ g^{0 }_\Lambda  G^{^1}_\Lambda
G^{^2}_\Lambda} ([A]) \ det_\Lambda^{^{1/2}}\Delta^{^0}_A\
 {\rm e}^{-S_{\Lambda}(A)}}
has a global meaning on the orbit space $\CM$ because volume element \seis\
under changes of
local coordinates   is similar to that of a
riemannian measure in  a finite dimensional manifold. Notice that
$\delta \mu_{ g^{0 }_\Lambda}=\delta  A_c  \ det^{^{1/2}}g^{\
}_\Lambda (A_c) $
 is the (formal) volume element of the
riemannian metric $g^{0}_\Lambda $, and the remaining  factors
 associated to the nuclear structures $G^{^1}_\Lambda$ and
$G^{^2}_\Lambda$ are gauge invariant.

\newsec{Finiteness and Gauge Invariance}

Let us analyze the finite character of the above
regularization in perturbation theory.
Because of the presence of a  large enough number of higher
covariant derivatives in the regulators of
the functional measure \cinco\ the degree of
ultraviolet divergences in diagrams involving more than one loops
is negative by power counting.
However, one loop divergences can not removed by this method and
have to be cancelled by an appropriate choice of the exponents of the
regulating terms.
The leading divergences arise in the 2-point gluonic function. They
 depend on the relative value of the leading exponents in the
scalar and axial terms $m$
and $n$ (see Appendix A for Feynman rules).

In order to properly handle one-loop divergences it is
convenient to introduce an auxiliary regularization because the
finite
radiative corrections might depend on the prescription used to
cancel the
divergences of each diagram.
A very natural
prescription
for the auxiliary ultraviolet regularization is the introduction
of a
momentum cut-off $\vert p\vert\leq \Omega$ for all propagating
modes.
The only problem with this prescription is that it breaks gauge
invariance
and is therefore necessary to impose some subsidiary conditions
to ensure
that it is restored when the pre-cutoff $\Omega$ is removed
\ref\rafe{M. Asorey, F. Falceto,
 Int. J. Mod. Phys. 7 (1992) 235-256}.

 The divergent contributions of gluonic loops come from the
diagrams (1) and (2) of fig. 1 at zero external momentum, and
are given by
\eqna\gcar
$$\eqalignno {&
  {2 h^{\vee}\over 3\pi^2}(m+1)^2\, \Omega\, \delta^{ab}\,\delta_{\nu\mu}
 &{\rm if} \quad m \geq 2n+1/2 \qquad\gcar{a} \cr
&  {2 h^{\vee}\over
3\pi^2}(2n+{3\over 2})^2\,\Omega\,\delta^{ab}\,\delta_{\nu\mu}
& {\rm if} \quad m \leq 2n+1/2 \qquad\gcar{b}
}$$
and
\eqna\ghierro
$$\eqalignno {
& -{2 h^{\vee}\over
3\pi^2} ((m+1)^2+{m\over 2})\, \Omega
\,\delta^{ab}\,\delta_{\nu\mu} & {\rm if}
\quad m\geq 2n+1/2 \ \ghierro{a}\cr
  &   -{2h^{\vee}\over
3\pi^2} ((2n+{3\over
2})^2+n+{1\over 4})
\,\Omega \,\delta^{ab}\,\delta_{\nu\mu}
&{\rm if} \quad m\leq
2n+1/2 \ \ghierro{b}
}$$
%
\vskip.2cm
{\hskip-.8cm \epsfxsize=14cm \epsfbox{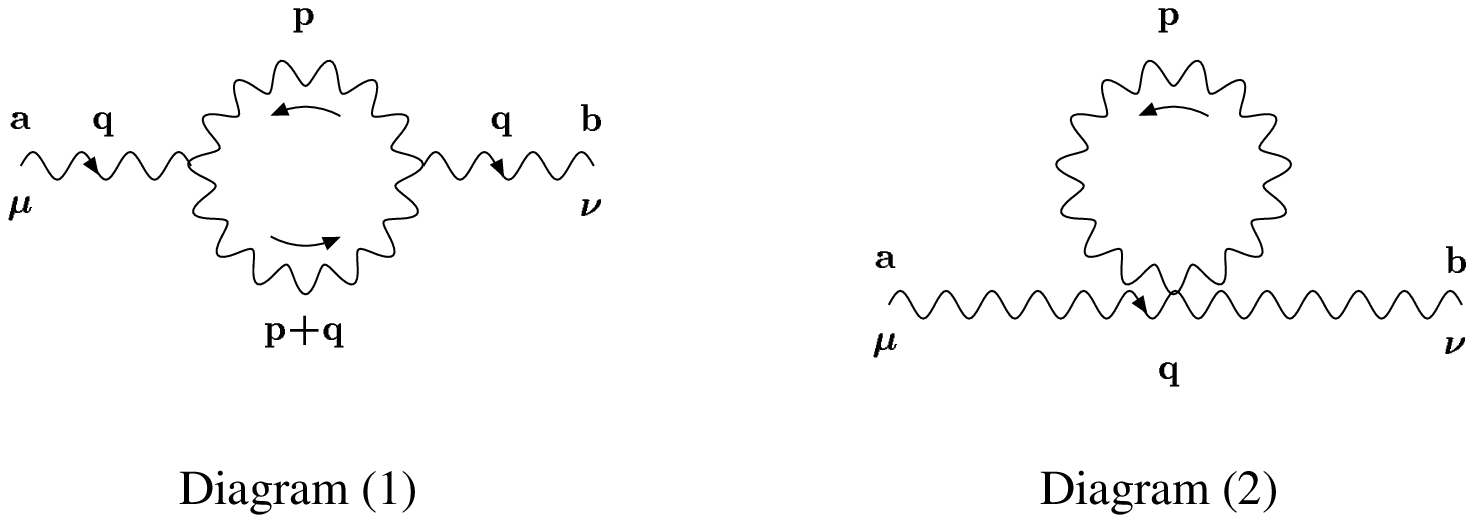}}
\vskip-.5cm
\centerline {{\bf Figure 1.} {\it \baselineskip=10pt plus 2pt minus 1pt
Radiative corrections to the vacuum polarization
involving gluon loops.}}\baselineskip=16pt plus 2pt minus 1pt
\medskip \noindent
respectively (see appendix B).
$h^{\vee}$ denotes the dual Coxeter number of $G$, e.g.
$h^{\vee}=N$
for $G=SU(N)$.
Therefore the total divergence associated to
gluonic loop reads
\eqna\gtot
$$\eqalignno { \quad  -&{h^{\vee}\over 3\pi^2}  m\,\Omega\,\delta^{ab}
\,\delta_{\nu\mu}
&{\rm if} \quad m\geq 2n+1/2.\qquad\gtot{a} \cr
 \quad -&{h^{\vee}\over 6\pi^2} (4n+1)\,\Omega\,\delta^{ab}\,
\delta_{\nu\mu}
&{\rm if} \quad m\leq 2n+1/2 \qquad\gtot{b} .\cr
}
$$
This divergent contribution to the vacuum polarization tensor can
be cancelled by the  contributions of
one-loop diagrams of metric and nuclear ghosts. The divergent
contributions generated by one loop radiative corrections  of
metric
ghosts (diagrams (1) and (2) of fig. 2) are
\vskip.2cm
{\hskip-.8cm \epsfxsize=14cm \epsfbox{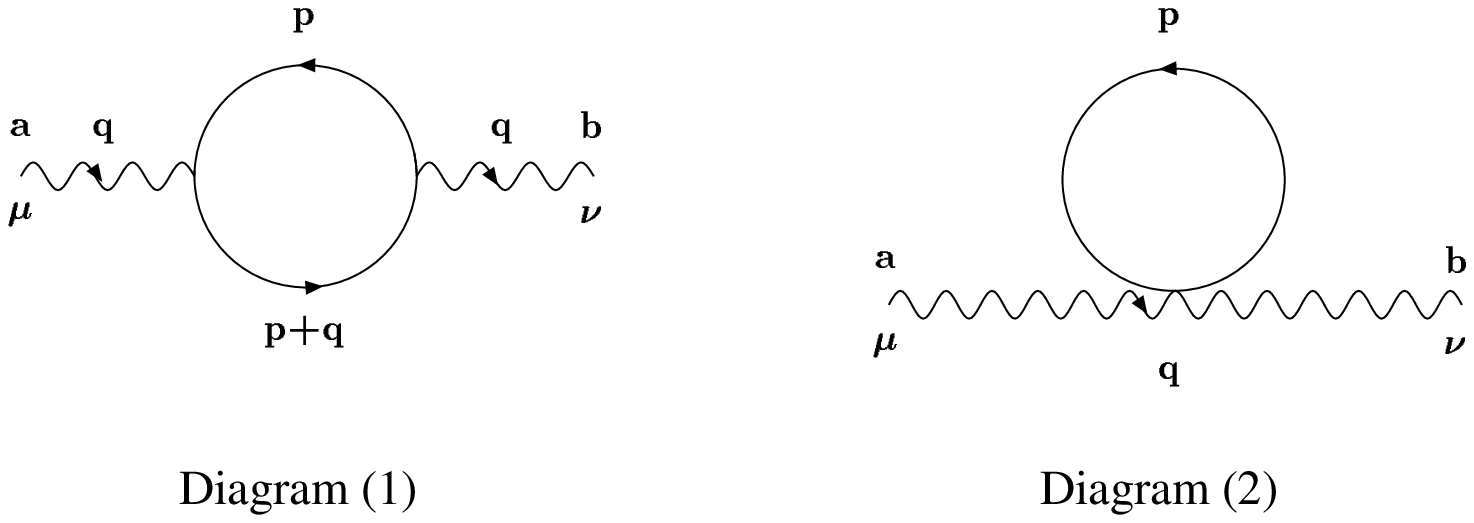}}
\vskip-.5cm
\centerline {{\bf Figure 2} {\it\baselineskip=10pt plus 2pt minus 1pt
 Radiative corrections to the vacuum polarization
involving ghosts loops. }}
\medskip\baselineskip=16pt plus 2pt minus 1pt
\eqn\eveintiuno{-{2h^{\vee}\over 3\pi^2}\, \Omega\delta^{a
b}\delta_{\mu\nu}{m_0^2}}
and
\eqn\eveintidos{{ h^{\vee}\over 3\pi^2}\left(2m_0^2 +{m_0}
-{1}\right)\,\Omega \delta^{ab}\delta_{\mu\nu},}
respectively.

In a similar way, the corresponding diagrams of nuclear ghost loops
yield
\eqn\eveintitres{-{2h^{\vee}\over 3\pi^2}
[(m_1-m_0)^2 +(m_2-m_1)^2]\,\Omega\, \delta^{a
b}\,\delta_{\mu\nu}}
and
\eqn\eveinticuatro{{h^{\vee}\over 3\pi^2}
\left({2(m_1-m_0)^2 }
+2(m_2-m_1)^2 +{m_2-m_0}\right)
\,\Omega \,\delta^{a
b}\, \delta_{\mu\nu},}
respectively.
Therefore the total divergent contribution of metric and nuclear
ghosts is
\eqn\eveintic{{h^{\vee}\over 3\pi^2}
{(m_2-1)}
\Omega\, \delta^{ab}\, \delta_{\mu\nu}.}
The sum of linear divergences \gtot{a}\gtot{b}\ and \eveintic\
vanishes provided
the finiteness condition
\eqn\ocho{m_2=\max\{m+1,2n+{3\over 2}\} }
is satisfied. The absence of  divergences in the radiative
corrections to the
ghost propagators is guaranteed by the choice of  regulating
exponents satisfying
the following inequalities
\eqn\ineq{\max\{m,2n+\ha\} > \{1,m_0,m_1-m_0,m_2-m_1\},}
Those conditions  are closely related to the
binuclear character of the orbit space $\CM$ in geometric
regularization and allow us to get rid of the problem of overlapping
divergences usually associated with regularizations by higher
covariant derivatives  \ref\FS{L.D. Faddeev, A. Slavnov,
{\it  Gauge Theories: Introduction to Quantum Theory},
Benjamin-Cummings, Reading (1980)}. The importance
of  nuclear structures can be understood from a perturbative
point of view as a necessary condition to compensate the ultraviolet
behavior of the ghost propagators with the singularities associated
to
ghost-gluon interactions.
The divergences
generated by the  scalar ghosts involved in the regularization of
the volume element of the gauge orbits  also cancel out under the
conditions
\ocho\ineq. In fact, those divergent contributions  are given by
$${4h^{\vee}\over 3\pi^2}
[m_0^2 + (m_1-m_0)^2 +(m_2-m_1)^2 - (m+1)^2]\,\Omega\, \delta^{a
b}\,\delta_{\mu\nu}$$
${\rm if} \quad m\geq 2n+1/2  $ or
$${4h^{\vee}\over 3\pi^2}
[m_0^2 + (m_1-m_0)^2 +(m_2-m_1)^2 - (2n+{3\over2})^2]\,\Omega\,
\delta^{a b}\,\delta_{\mu\nu}$$
${\rm if} \quad m\leq 2n+1/2$
for diagrams of type (1), and
$$-{2h^{\vee}\over 6\pi^2}
\left(4m_0^2+{4(m_1-m_0)^2 }
+4(m_2-m_1)^2 +{m_2}-4(m+1)^2-m-1\right)
\,\Omega \,\delta^{a
b}\, \delta_{\mu\nu}$$
${\rm if} \quad m\geq 2n+1/2 $ or
$$-{2h^{\vee}\over 6\pi^2}
\left(4m_0^2+{4(m_1-m_0)^2 }
+4(m_2-m_1)^2
+{m_2}-4(2n+{3\over2})^2-2n-{3\over2}\right) \,\Omega
\,\delta^{a b}\, \delta_{\mu\nu}$$
${\rm if} \quad m\leq 2n+1/2$
for diagrams of type (2). They
cancel  each other out under the condition \ocho.
The splitting of the
cancellation of one loop
divergences into two parts, one corresponding to vector fields and
another to scalar fields, is one of the  characteristics of geometric
regularization. Divergences generated by one gluonic loop diagrams
are cancelled by
the  diagrams associated
to  the volume element of the binuclear structure of the orbit space
and  the divergences
corresponding to the volume of the gauge fibers are regularized by
standard Pauli-Villars methods.

  On the other hand there are not divergences associated to
three-point functions. In this case the potentially logarithmic
divergent terms
cancel out by algebraic reasons and the final contribution always
remains  finite.

Although the regularization is formally gauge invariant the
introduction of the auxiliary momentum cut-off $\Omega$ for the
analysis of  the cancellation of one loop divergences breaks
explicitly gauge invariance.
This invariance might be recovered after removing  $\Omega$,
but in general the restoration of the symmetry requires to
impose some constraints on the exponents of the regularization \rafe.
 The
conditions  which guarantee the  vanishing of anomalous terms in
Slavnov-Taylor  identities are
\eqn\gi{\max\{m+1,2n+{3\over 2}\}^2= m_0^2+(m_1-m_0)^2+(m_2-m_1)^2.}
The constraints arise from the analysis of  possible sources of
violation of those identities. The diagrammatic derivation of
the identities involves a comparison of the contributions of
diagrams of type (1) and (2) in  two points functions. Now,
the domain of integration in diagrams of type (1) is
$\vert p\vert\leq\Omega$, $\vert p+q\vert\leq\Omega$ whereas in
diagrams of type (2) it is just $\vert p\vert\leq\Omega$, and
such a difference
of domains generates an  anomalous contribution
because both diagrams were originally divergent.
Three and
higher point functions, being  logarithmically divergent or finite,
do not generate in the limit of
$\Omega\rightarrow\infty$ any  anomalous contribution associated
to the differences of domains of integration in diagrams
with different topologies.

A way of overcoming the problems with the choice of
domains in Feynman integrals
is to impose finiteness on the sum of diagrams with the same
topology separately,  then the sum of the
 integrals associated to diagrams of type (1) is
  finite when $\Omega\to\infty$  and it is independent of
which of the two domains of integration we consider.

It can be shown that it is precisely the condition \gi\
which together with \ocho\ implies the vanishing of the
sum of divergences of  diagrams with
the same topology.  In particular, the divergent contributions of
diagrams of type (1) of figs. 1 and 2, \gcar{a}\gcar{b},
\eveintiuno\ and \eveintitres\ cancel out if and only if \gi\ is
satisfied. The cancellation of the contributions of diagrams of
type (2) follows from \ocho\ and \gi.

If  conditions \ocho\ and \gi are satisfied
we can completely forget about the  auxiliary cut-off $\Omega$
and use the standard Pauli-Villars prescription where the internal
momenta of  all divergent diagrams with the same topology are
parametrized in an identical way. However the use of the auxiliary
cut-off $\Omega$ is very convenient because  it provides a
non-perturbative approach to the functional integral and
gives an unambiguous prescription for the parametrization of
individual diagrams in Pauli-Villars regularizations which
involve different fields with different gauge interactions.

After this short digression regarding the relevance of \ocho\ and
\gi\ to guarantee finiteness and gauge invariance in the
regularization, we shall consider the solutions of those conditions.
 When the scalar part of the action dominates the
ultraviolet behavior the general solution of conditions \ocho\ and
\gi\ with integer exponents  is  \rafe\
\eqnn\ediecicocho
$$\eqalignno {& m+1= m_2 = c\, (a^2+b^2+ab)\cr
& m_0 = c\, (a^2+ab) &\ediecicocho \cr
& m_1 = c\, (a^2+b^2+2ab) \cr}$$
where
 $a,b,c$ are any three positive integers
with $a,b$ relatively primes and $a<b$.
 When the pseudoscalar part of the action is ultraviolet
dominant there are not solutions with integer coefficients. This
implies the presence of non-local pseudo-differential operators in the
regulating terms of ghost-gluon interactions. An explicit solution
of the conditions in this case
can be given by,
\eqnn\edieciocho
$$\eqalignno {& 4n+3=2 m_2 = c\, (a^2+b^2+ab)\cr
&2 m_0 = c\, (a^2+ab) &\edieciocho\cr
&2 m_1 = c\, (a^2+b^2+2ab), \cr}$$
as in \ediecicocho.
An infinity of values of $a,b$ and $c$ satisfy the
remaining inequality constraints \ineq.

\newsec{One-loop Radiative Corrections}

 Once we have shown that the regularization cancels all
perturbative divergences we can address the calculation of finite
corrections  generated by one loop diagrams.
The finite scalar contributions to the two-point function
generated by one gluon loops are
\eqn\gluonss{
-{h^{\vee}\over 8\pi^2}\, (m+1)^2 \vert q \vert\left (
\delta_{\mu\nu} +
{q_{\mu}q_{\nu} \over q^2} \right )\, \delta^{ab}  +
\CO(\Omega^{-1},\Lambda^{-1})}
${\rm if} \quad m\geq 2n+1/2 $ and
\eqn\gluonsss{-{h^{\vee}\over 8\pi^2}\, (2n+3/2)^2 \vert q
\vert\left (  \delta_{\mu\nu} +
{q_{\mu}q_{\nu} \over q^2} \right )\, \delta^{ab}  +
\CO(\Omega^{-1},\Lambda^{-1})}
${\rm if} \quad m\leq
2n+1/2 $ (see Appendix B).
These contributions cancel up to terms of order
$\CO(\Omega^{-1},\Lambda^{-1})$  with those generated
by metric and
nuclear ghost loops
\eqnn\gluons
$$\eqalignno {&{h^{\vee}\over 8\pi^2}\, (m_0)^2 \vert q \vert\left (
\delta_{\mu\nu} +
{q_{\mu}q_{\nu} \over q^2} \right )\, \delta^{ab}  +
\CO(\Omega^{-1},\Lambda^{-1}) \cr
&{h^{\vee}\over 8\pi^2}\, (m_1-m_0)^2 \vert q \vert\left (
\delta_{\mu\nu} +
{q_{\mu}q_{\nu} \over q^2} \right )\, \delta^{ab}  +
\CO(\Omega^{-1},\Lambda^{-1}) &\gluons \cr
&{h^{\vee}\over 8\pi^2}\, (m_2-m_1)^2 \vert q \vert\left (
\delta_{\mu\nu} +
{q_{\mu}q_{\nu} \over q^2} \right )\, \delta^{ab}  +
\CO(\Omega^{-1},\Lambda^{-1}) \cr
}
$$
by  condition \gi. Similar contributions are generated by ghosts
associated to the volume of the gauge orbits \ecinco\ but cancel
out among
themselves by the same reason. Non-trivial contributions to the
mass
term generated by diagrams of type (1) cancel out with those
of diagrams of type (2).
On the other hand it is easy to see that the anomalous
contributions
to the Slavnov-Taylor identities involving two and three point
functions
also cancel out when the condition \gi\ is satisfied. The
non-analytic dependence of  expressions \gluonss\ --\gluons\ come
from the constraint imposed by the non-perturbative cut-off $\Omega$
on the internal momenta of diagrams of type (1),
$|p|<\Omega$ and
$|p+q|<\Omega$.

After the
removal of the ultraviolet regulating parameter $\Lambda$ the only
quantum  corrections to the two and three point functions are local
and pseudoscalar like the two terms of the Chern-Simons action.
	The radiative corrections to the two point Green function
generated by gluonic loops (diagrams (1)--(2) of fig. 1) are given
by
\eqn\snine{ ^{}\Gamma_{\mu\nu}^{a b}={h^{\vee}\over3\pi^2}
 \int_{0}^\infty dp\, {\Sigma(p) \over
\rho(p)}\delta^{a
b}\epsilon_{\mu\sigma\nu} q^\sigma,}
where
\eqn\sten{\eqalign{
\Sigma(p)=&2 S(p)+{3 \over2} \bigg\lbrack R(p)
{d(S(p)p)\over dp}-pS(p) {dR(p)\over dp} \bigg\rbrack
,\cr}}
\eqn\sthree{\rho(p)= p^2 S(p)^2+
R(p)^2,
}
and
$$
S(p) = \lambda (1+p^2)^m \qquad
R(p) = 1+\lambda' p^2(1+p^2)^{2n}
$$
(see appendix B for more details).

There are not further one loop corrections to the pseudoscalar part
of the  effective
action because loops of metric and nuclear ghosts do not
involve pseudoscalar couplings.

The radiative correction to the quadratic term of the pseudoscalar
effective action contains the renormalization of the gauge field and
the renormalization of the coupling constant. In order to split  the
two contributions we calculate the renormalization of the
gauge field by analysing the renormalization of the coupling of
gauge fields to Faddeev-Popov ghost\foot{Although those ghosts do
not appear in the geometric formulation of the functional integral
they appear in  Slavnov-Taylor identities which are also
satisfied  by the
regularization provided the condition \gi\ is satisfied \rafe.}.
The renormalization of the
ghost field $c$ is obtained from the diagram (1) of fig. {3}
whose contribution reads
\eqn\sss{\Sigma_{_{a b}}=q^2{4 h^{\vee}\over
3k\pi}\int_{0}^\infty {S(p)\over \rho(p)}dp\ \delta_{a b}.}
This leads to a finite renormalization of Faddeev-Popov ghost fields
\eqn\renos{\ c_R= \bigg[1-{2 h^{\vee}\over
3k\pi}\int_{0}^\infty {S(p)\over \rho(p)}dp\bigg] \ c\  }
One-loop radiative corrections to the gluon-ghost interaction
(diagrams (2) (3) of fig. 3)
vanish. Therefore the renormalization of the gauge field is given
by
\vskip 4cm
{\hskip-.8cm \epsfxsize=14cm \epsfbox{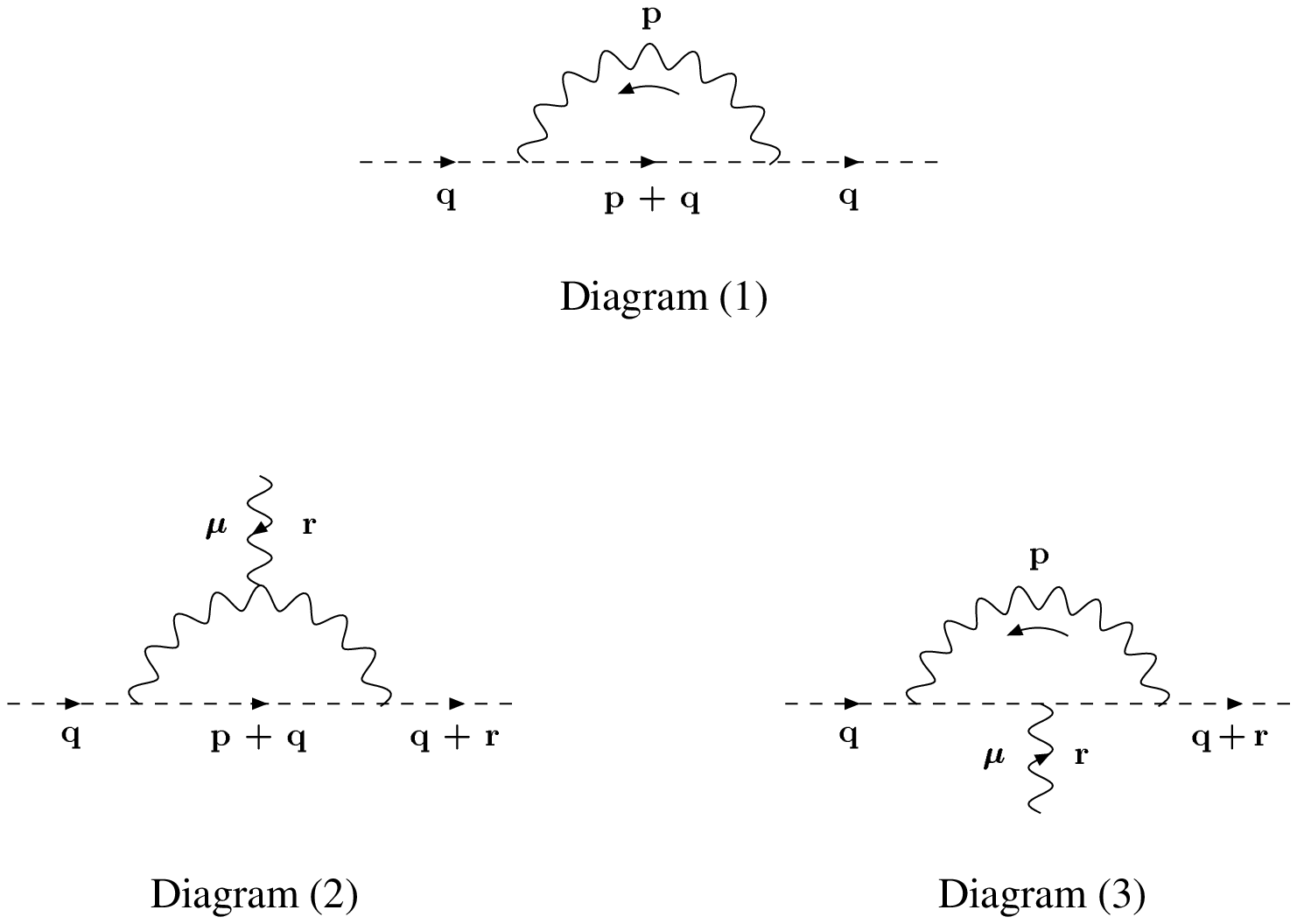}}
\vskip-.0cm
\baselineskip=10pt plus 2pt minus 1pt
\noindent {{\bf Figure 3} {\it 
  One loop contributions to the   self-energy of the
Faddeev-Popov ghosts (diagram (1)), and
  to the
3-vertex gluon-ghost interaction (diagrams (2) and (3)).}}
\medskip\baselineskip=16pt plus 2pt minus 1pt
\eqn\ren{A_\mu^R=\bigg\lbrack 1 +{4 h^{\vee}\over
3k\pi}\int_{0}^\infty {S(p)\over \rho(p)}dp\bigg\rbrack\  A_\mu.}
We remark the existence of a non-trivial renormalization of the gauge
field $A$ except in the pure axial case $\lambda=0$.

 Plugging the
renormalized fields into the two point gluonic function yields a
finite renormalization of the \cs\ coupling constant of the form
\eqn\shift{k_R=k+ {4h^{\vee} \over 3\pi}\left\lbrack
\int_{0}^\infty {\Sigma'(p)
\over \rho(p)} dp\right\rbrack
}
where
$$
\Sigma'(p)= \Sigma-2S(p).
$$
If we perform the change of variables $\phi=\phi(p)=pS(p)/ R(p)$
the  second term
on the right hand side of \shift\
for $\lambda' > 0$ becomes
\eqn\rot{{2h^{\vee}\over\pi}
\int_{\phi(0)}^{\phi(\infty)}
{
d\phi \over 1+\phi^2}, }
%
%
%
 and  can be exactly calculated. However, when $\lambda'<0$ the change of
variables has a singular point $ p_\infty$ where
$\phi(p_\infty)_{\pm}=\mp \infty$ (see fig. 4).
\vskip.2cm
{\hskip-.8cm \epsfxsize=14cm \epsfbox{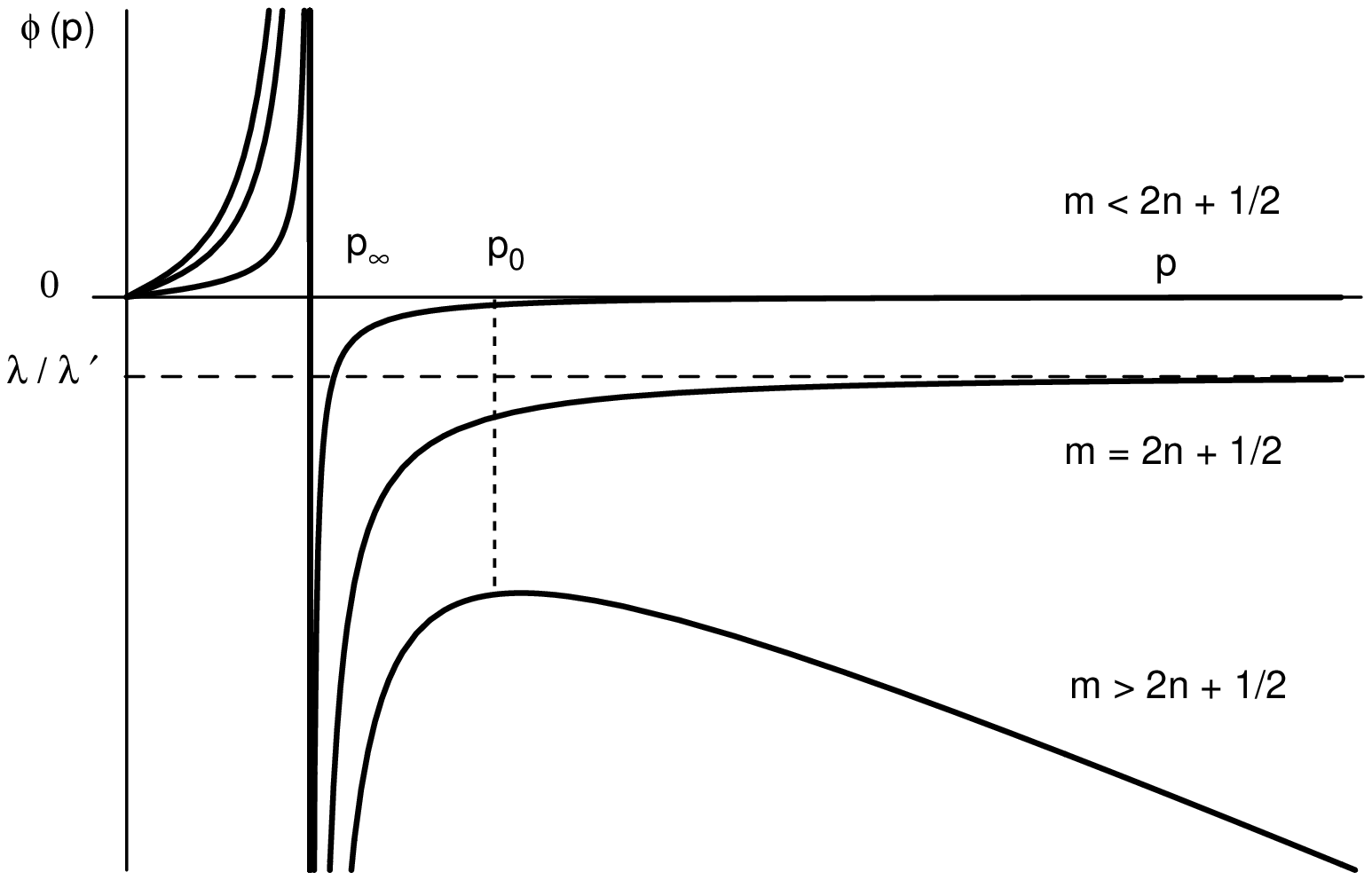}}
\vskip-.0cm
 \baselineskip=10pt plus 2pt minus 1pt
\noindent {{\bf Figure 4} {\it
 { The three regimes of ultraviolet behavior of
the $\phi$ function for $\lambda'<0$. The infinite gap at $p_\infty$
corresponds to a zero value of the pseudoscalar leading
term which appears in all  regimes. }}}
\medskip\baselineskip=16pt plus 2pt minus 1pt
Therefore, the domain of
integration in  $\phi$ variable  splits into
two domains
\eqn\rott{{2h^{\vee}\over\pi}
\left\lbrack\int_{\phi(0)}^{\phi(p_\infty)_-}
+\int_{\phi(p_\infty)_+}^{\phi(\infty)}\right\rbrack{
d\phi \over 1+\phi^2}. }

There are three different regimes  for 
the behavior of the
effective  coupling constant  $k$  
(see Figs. 4,5)
\foot{An additional shift by $2sh^\vee$ could
be obtained in any regime by considering pseudoscalar couplings
between gluons and ghosts \rlc. For simplicity, we shall not
consider such a possibility in this paper. We will restrict
ourselves to the analysis of the behavior of the  corrections
generated only by gluon selfinteractions.}:
\vskip-.5cm
{\hskip-.8cm \epsfxsize=14cm \epsfbox{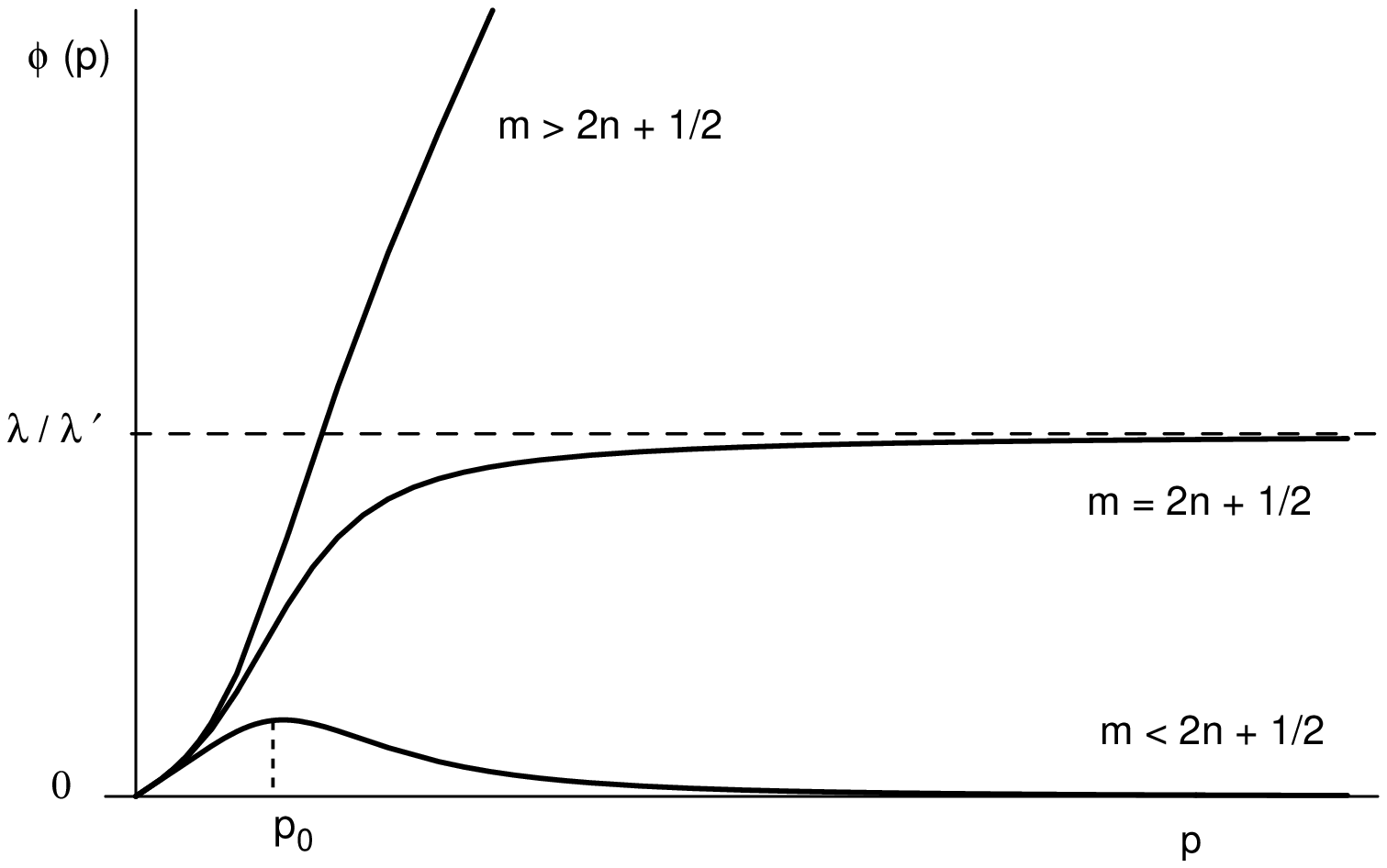}}
\vskip-.0cm
\baselineskip=10pt plus 2pt minus 1pt
\noindent {{\bf Figure 5.} {\it 
  { The three regimes of ultraviolet behavior of the
$\phi$ function for $\lambda'>0$. $p_0$ is the critical
point where
the function $\phi$ reachs its maximal value in the regime
$m<2n+{1\over 2}$.}}}

\medskip\baselineskip=16pt plus 2pt minus 1pt

 i) If $m>2n+1/2$ and   $\lambda' > 0$,
since $\phi(p)$ is a   continuous function with $\phi(0)=0$ and
$\phi(\infty)=\infty$ the shift is universal and independent of the
parameters of the regularization
\eqn\cuno{k_R= k + h^{\vee}}
%
The same shift occurs in that case  for $\lambda' < 0$
because the second integral in \rott\ vanishes.
This regime includes all the cases where the ultraviolet divergences
are regulated by Yang-Mills like terms \pisrao--\carmelo.  Most
of the regularizations analysed in the literature are of this  type.

ii) If $m<2n+1/2$,   $\phi(0)=0$
and $\phi(\infty)=0$. Therefore, if $\lambda'>0$, $\phi(p)$ is a
continuous function and there is no shift of $k$ for any value  of the
other parameters of the regularization.
This explains why it is possible to find a gauge
invariant regularization scheme  where the  effective value of
the coupling constant equals its bare value \ita.
 This  universality class of regularizations
was first analyzed in a gauge invariant way in ref.
\rlc. However there is another universality class associated to
the same regime. If $\lambda'<0$ the function $\phi(p)$ develops
 a pole at the only real root $p_\infty$ of the polynomial
$1+\lambda'p^2(1+p^2)^{2n}$. Therefore the shift of $k$ is in
this case $k+2h^\vee$. This gives rise to a new universality
class valid for all regularizations with leading pseudoscalar
coupling with opposite sign to the Chern-Simons term.

iii) If $m=2n+1/2$, since $\phi(0)=0$ but $\phi(\infty)=
\lambda/\lambda'$
the shift is not universal and does depend on the parameters
of the
regularization
\eqn\cuno{k_R= k + {2h^{\vee}\over\pi}
\arctan{\lambda\over\lambda'},}
with $\arctan$ taking values in $[0,\pi)$.
This is a novel regime and its discovery is one of the major
issues of this
work.

One remarkable aspect of the above results is
the existence of  universality
in the
renormalization of $k$ in the first two regimes and its violation
in the borderline case. The physical explanation of this fact can
be learned from  \rot\ and \rott\ where it is shown that the shift
is  originated by the balance between
ultraviolet and infrared  behaviors of the regulating terms of the
real  and imaginary parts of the  regularized action.
The infrared behavior is unchanged in  the three different cases
whereas the ultraviolet regime depends on which terms has more
derivatives. Once there exists a dominant scalar or axial part the
shift is
constant and therefore universal; at most it might depend on the
sign of the leading pseudoscalar term $\lambda'$. However, in the
borderline case with a  hybrid ultraviolet behavior the result
depends  on the relative weight of  the
coefficients of the terms with higher number of derivatives.
In
such a case  any value for the effective
coupling constant can be attained.

Notice that
the constraint associated to the invariance under
large gauge transformations only applies to the bare coupling
constant.
Therefore nothing prevents the existence of non-integer values
for the effective coupling constant. The reason why this
possibility
has never considered previously is because it required a
regularization
by means of pseudodifferential operators \foot{ Notice that one
of the two
exponents $m$ or $n$
have to be half integer in the borderline case
because of the identity $m=2n+1/2$}.
Any other choice leads to an
integer valued effective coupling constant \foot{ In fact, in the
three regimes above discussed there is a larger ambiguity in the
shift induced by radiative corrections. Besides the shift by $2s
h^\vee$ induced by radiative corrections generated by Pauli-Villars
ghost fields with pseudoscalar couplings afore mentioned, there is
another source of ambiguity of the same type purely generated by
gluon radiative corrections. If instead of regulating the
Chern-Simons action by polynomials on the covariant laplacian
$\Delta_{A}$ of the
form $(1+\Delta_{A})^n$ we consider a more general polynomial
in $\Delta_{A}$ the
$\phi(p)$ function might develop a  richer critical structure  with more
zeros and poles. In particular, if it has $s$ poles with $s-1$ zeros
intercalated between those poles the integrals \rot\ and \rott\ would
give rise
to an additional shift of $k$ by $2sh^\vee$ for any of the three
regimes discussed above.}.

In the case where the effective coupling constant is non-integer the
connection with conformal field theory is presumable through
non-rational conformal field theories.

 \newsec{Topological Anomaly}

In order to explore further consequences of the existence of three
renormalization regimes of the coupling constant we investigate the
behavior of the radiative corrections to the vacuum energy to
elucidate whether the three
regimes give rise to different physical effects in this case too.

In particular, since Chern-Simons theory is a topological theory
the radiative corrections should preserve the vanishing of the
vacuum
energy. However due to the existence of a topological anomaly a
finite
contribution can appear in the form of a gravitational Chern-Simons
term
\eqn\gcs{G_{\rm cs}={i \kappa\over 4\pi }\int\
\tr
\left(\omega\wedge d\omega+ {2\over 3}
\omega\wedge\omega\wedge\omega
\right).}
defined by the Levi-Civita connection $\omega$ of $TM$
and a non-local framing-dependent term (framing anomaly)
with a non-analytic dependence on the dreivein field $e_\mu^a$
\ref\witten{E. Witten,
Commun. Math. Phys. 121 (1989) 351-399
}.

If the space time
manifold is of the form $\Sigma\times S^1$ and the metric
splits into a product of metrics of  $\Sigma$ and  $S^1$,
there exists
a framing  where the gravitational Chern-Simons term vanishes.
Therefore the  partition  function
becomes completely metric independent. However, for a general three
manifold $M$ and arbitrary framing of its tangent bundle $TM$
the gravitational Chern-Simons terms does not vanish
(see \ref\ind{M. Asorey,
F. Falceto,  J.L. L\'opez and G. Luz\'{o}n,
In preparation}
for a general discussion of metric
(in)dependence of the partition function of  Chern-Simons theory).
Moreover
 Witten conjectured an universal behavior for the coefficient of
this
pseudoscalar term of the form \witten\
$$\kappa={c\over 24} $$
which provides a three-dimensional interpretations of the
 central charge of the corresponding Wess-Zumino-Witten
model,
\eqn\cch{c= {k \dim G\over k + h^{\vee}}.}

Since $c$ is an analytic function of ${h^{\vee}/ k}$
\eqn\pert{c= \dim G \sum_{n=0}^{\infty}
\left({-h^{\vee}\over k}\right)^{n}.
}
it can be computed in perturbation theory.

We shall calculate the first corrections to this coefficient in the
regularization scheme discussed in the previous sections to see if
the
existence of three renormalization regimes has some consequences on
Witten's
conjecture.

The remaining non-local contributions can also be
indirectly calculated in this approximation and its presence is
reflected in the existence of  a framing
anomaly. It follows from the
observation that since the regularization method is
independent of any framing of the tangent bundle. Then the
effective action must be framing independent. Now the gravitational
Chern-Simons term is frame dependent, therefore, it must exist a
non-analytic locally constant
dependence on the dreivein field $e^i_{\alpha}$ whose
variation under large frame transformations counterbalance the
variation of the gravitational
Chern-Simons term \foot{Those non-analytic
terms also appear in the spectral asymmetry function $\eta$
which is associated  \witten\ref\piet
{L. Alv\'arez-Gaum\'e, S. Della Pietra, G. Moore, Ann. Phys.
{  163} (1985) 288} to the elliptic operator $\ast d+ d\ast$ acting
on Lie--$G$ valued one-forms and are connected with its spectral
flow and the index of its four dimensional extension.}. We can eliminate
the contribution of the Chern-Simons gravitational term by a local
counterterm  but the  contribution of the non-local term cannot be
cancelled
 in such a way and this fact implies the existence of a
framing anomaly in such a case \foot{However, it is always possible
to choose a framing of the double-tangent bundle $T^2M$ of $M$ (the
canonical framing) where this non-local framing dependent term
vanishes \ref\freed{D.S.  Freed, R.E. Gompf, Commun. Math. Phys. {
141}(1991)79}\ref\atiyah{M. F.  Atiyah, Topology { 29}(1990) 1}}.

A general feature of the geometric regularization method is
that once the ultraviolet divergences are cancelled in a flat
space-time
they do not reappear when the theory  is defined on
curved background metric \ref\ind{M. Asorey,
F. Falceto,  J.L. L\'opez and G. Luz\'{o}n,
Proceedings of the  XIX International Colloquium
on Group
Theoretical Methods in Physics, M. del Olmo et {\it al.} eds.,
Ciemat, Madrid (1993) 55}.
 \nref\axel{ S. Axelrod, I.M. Singer,  Proceedings
of the XXth
International Conference on Differential Geometrical Methods
in Theoretical
Physics, S. Catto and A. Rocha eds., World Sci. (1991)}
\nref\pig{C. Lucchesi, O.
Piguet, Nucl. Phys. { B381}(1992) 281} However,
when the regulator is removed some
divergences might appear  unless we take special care of the
Pauli-Villars regularization of the  scalar ghost fields associated
to the volume of the orbits of the gauge  group. The special
form of this regularization chosen in \ecinco\ guarantees that
not only
the theory is  completely regularized in our case, but  there
is not need of
an infinite renormalization in  curved spacetimes backgrounds,
at least at
one loop level.
 A general discussion of  finiteness and metric
(in)dependence of the partition function on arbitrary curved
space-times
can be found in \ind-\pig.

In this section we are only interested
on the
calculation of local (finite) radiative corrections  of pseudoscalar
type in order to verify Witten's conjecture.

Since  ghosts fields have only scalar interactions the lowest
order pseudoscalar
radiative corrections can only  appear  from one gluon loop
corrections
to the vacuum energy,
\eqn\ongl{S_{\rm eff}= -\half\log \det_c (({\lambda}(I+\Delta/
\Lambda^2)^m d^\ast d/\Lambda- i \ast d\left(I+ \lambda'(I+\Delta/
\Lambda^2)^{2n}d^\ast d/
\Lambda^2  \right).}
Due to the independence on the gauge field $A$, the  calculation
of the
determinants simplifies considerably  in operator formalism.
 In particular the transverse projectors
necessary to restrict the differential operators to
transverse modes
can be suppressed because all the operators are already
transverse, i.e.
\eqn\efaction{S_{\rm eff}= -\half \tr_c\log  (({\lambda}
(I+d^\ast d/
\Lambda^2)^m d^\ast d/\Lambda- i \ast d\left(I+
\lambda'(I+d^\ast d/
\Lambda^2)^{2n}d^\ast d/
\Lambda^2 \right).}
The pseudoscalar terms  come from the imaginary part of the
effective
action
\eqn\im{{i\over 2} \tr_c\arctan{\ast d\left(I+ \lambda'(I+d^\ast d/
\Lambda^2)^{2n}d^\ast d/
\Lambda^2 \right)\over{\lambda}(I+d^\ast d/
\Lambda^2)^m d^\ast d/\Lambda }.}
Now, since the gravitational Chern-Simons term is analytic in
metric
variables $g_{\mu\nu}$ we can calculate its coefficient by a weak
metric
expansion.
If we consider a background metric
$g_{\mu\nu}=\delta_{\mu\nu}+h_{\mu\nu} $ in $\IR^3$ which is a
slight
perturbation of the Euclidean metric $\delta_{\mu\nu}$,
the first contribution to the \cs\ gravitational term is
quadratic in the
metric perturbation $h$ and
cubic in space-time derivatives
\eqn\gcse{G_{\rm cs}={i \kappa\over 8\pi }\int\ d^3 x
\epsilon^{\mu\sigma\nu}
\left(\partial_{\nu'}h_{\mu\mu'}-\partial_{\mu'}h_{\mu\nu'} \right)
\partial_{\sigma}
\left(\partial_{\mu'}h_{\nu\nu'}-\partial_{\nu'}h_{\nu\mu'} \right)
+\CO(h^3)
.}
We remark that the expression \im\ is a functional of the single
operator
$\ast d$. Therefore, its
second variation with respect to the metric perturbation yields,
\eqn\var{\eqalign{{i\over 2}{\delta^2\over \delta g_{\mu\nu}
\delta g_{\mu'\nu'}}&
\tr_c\arctan{\ast d\left(I+ \lambda'(I+d^\ast d/
\Lambda^2)^{2n}d^\ast d/
\Lambda^2 \right)\over{\lambda}(I+d^\ast d/
\Lambda^2)^m d^\ast d/\Lambda }\cr=
& {i\over 2}\sum^\infty_{j=1}a_j
{\delta^2\over \delta g_{\mu\nu}\delta
g_{\mu'\nu'}}\tr_c (\ast d)^{2j+1}\cr}}
where $a_j$ are the coefficients of the Taylor expansion of the
analytic function
\eqn\tay{\Xi(t)= \arctan{\left(1+\lambda'(1+t^2/
\Lambda^2)^{2n}t^2/\Lambda^2  \right)\over{\lambda }(1+t^2/
\Lambda^2)^m t/\Lambda}= \sum^\infty_{j=0}a_j t^{2j+1}}
We only need to recall the expression
of the differential operator $\ast d$ in local coordinates,
\eqn\star{(\ast d)_\mu^\nu= {1\over \sqrt{g}}g_{\mu\gamma }
\epsilon^{\gamma\sigma\nu}\partial_\sigma.}
Hence,
\eqn\starr{\left.{\delta (\ast d)_\mu^\nu\over
\delta g_{\alpha\beta}}
\right|_{\hat{g}}=-\half {1\over \sqrt{\hat g}}
\hat {g}_{\mu\gamma}
\hat  {g}^{\alpha\beta}
\epsilon^{\gamma\sigma\nu}\partial_\sigma+
{1\over \sqrt{\hat g}} {\delta}_{\mu}^{\alpha}
{\delta}^{\beta}_\gamma
\epsilon^{\gamma\sigma\nu}\partial_\sigma}
which  in our particular case of flat background metric $\hat
{g}_{\mu\nu}=
{\delta}_{\mu\nu}$ reads,
\eqn\starrr{{\delta (\ast d)_\mu^\nu\over \delta
g_{\alpha\beta}} =-\half \delta_{\mu\gamma}{\delta}^{\alpha\beta}
\epsilon^{\gamma\sigma\nu}\partial_\sigma+
\delta_{\mu}^{\alpha}\delta^{\beta}_{\gamma}
\epsilon^{\gamma\sigma\nu}\partial_\sigma.}
Now, terms which involve traces of $h_{\alpha\beta}$ do not
appear in the gravitational \cs\ action $G_{\rm cs}$.
Therefore, the first
term of
\starr\  do not contribute to $G_{\rm cs}$. The second term is not
dependent
on $\hat{g}$, which implies that the relevant dependence of $\ast d$
on the metric $\hat{g}$ is linear. Hence the
non-trivial  contribution of \var\ to $G_{\rm cs}$ is
\eqn\rosca{
{i\over 2} \sum^\infty_{j=1}\sum^{2j-1}_{l=0}
(2j+1) a_j \tr{\delta (\ast d)\over \delta g_{\mu\nu}}(\ast d)^{l}
{\delta (\ast d)\over \delta g_{\mu'\nu'}}(\ast d)^{2j-l-1}.}
Since we are perturbing the Euclidean metric of $\IR^3$,
 the trace
\eqn\trac{\tr_c{\delta \ast d\over \delta g_{\mu\nu}}(\ast d)^{l}
{\delta \ast d\over \delta g_{\mu'\nu'}}(\ast d)^{2j-l-1}}
can be evaluated in momentum space.
The contribution of the one loop diagrams to
the quadratic graviton-graviton pseudoscalar term \var\
reads (see fig. 6),
\vskip.2cm
{\hskip-.8cm \epsfxsize=14cm \epsfbox{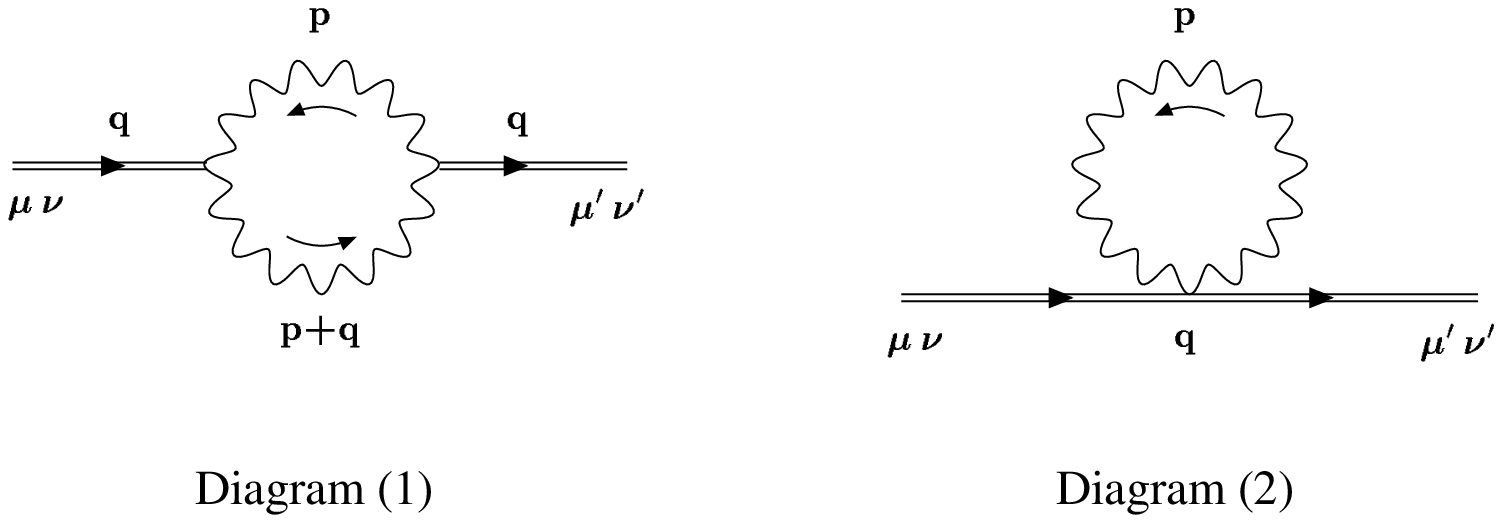}}
\vskip-.5cm
\centerline{{\bf Figure 6.}{\it One loop   contributions to the 
graviton 2-point function involving gluon loops.}}
\medskip
\eqn\ggrr{\eqalign{
-\half \dim G &\int {d^3 p\over (2\pi)^3 }
\left(\epsilon^{\nu\eta\mu'}
(p+q)_\eta
\left[ p^2\delta_{\nu'}^{\mu}-p_{\nu'} p^\mu\right]\right.
\cr
& +\epsilon^{\nu'\eta\mu}
p_\eta\,
\left[ \left. (p+q)^2\delta_{\nu}^{ \mu'}-(p+q)_{\nu}
(p+q)^{\mu'}\right]\,\right)\cr &
\sum^\infty_{j=1}(2j+1) a_j
\sum^{j-1}_{s=0}
(p+q)^{2s}
p^{2j-2s-2}\cr}}
where $q$ denotes the external momentum.
In order to get the local terms with cubic
dependence on the external momentum we have to expand the function
\eqn\fun{\Theta(p,q)=\sum^\infty_{j=1} (2j+1) a_j
\sum^{j-1}_{s=0}(p+q)^{2s}
p^{2j-2s-2}={\left[{ d\, \Xi(p+q)\over
d p }-{ d\, \Xi(p)\over
d p }\right]\over  (p+q)^2-p^2}}
up to third order in the power Taylor expansion.
Plugging the corresponding terms
\eqn\tayl{\eqalign{\Theta(p,q) &= {1\over 2p} {d^2\, \Xi(p)\over
d p^2 }+{1\over 4} (p.q) \left({1\over p}{d \over
d p }\right)^2{ d\, \Xi(p)\over
d p }+{1\over 12} (p.q)^2 \left({1\over p}{d \over
d p }\right)^3{ d\, \Xi(p)\over
d p }\cr
+ &
{1\over 8}q^2\left({1\over p}{d \over
d p }\right)^2{ d\, \Xi(p)\over
d p }+{1\over 48} (p.q)^3 \left({1\over p}{d \over
d p }\right)^4{ d\, \Xi(p)\over
d p }\cr
+ &
{1\over 12} (p.q)q^2 \left({1\over p}{d \over
d p }\right)^3{ d\, \Xi(p)\over
d p } + \CO(q^4)\cr
}}
into \ggrr, using  symmetry properties of
 metric tensors and
internal momentum integrals and integrating by parts we get
\eqn\cond{- {1\over 48\pi^2}\dim G\epsilon^{\mu\sigma\mu'}
q_\sigma(q^2 \delta_{\nu\nu'}- q_{\nu}q_{\nu'})\int_0^\infty
dp {d\,
\Xi(p)\over dp}.}
In fact, only the second and third terms of the $q$--expansion
\tayl\ give a non trivial contribution because the other
terms cancel out. The $q$--independent  and $q$--linear
contributions which are potentially linear and logarithmic
divergent
in the borderline case also vanish
\foot{In this case one must use
the Feynman rules for pseudodifferential
operators. We follow the standard prescription in analogy
to the Feynman rules for  gauge field
couplings described in Appendix A }.

This implies that the quantum corrections to the vacuum energy
generate a gravitational \cs\ term $G_{\rm cs}$ with coefficient
\eqn\will{\kappa=-
{\dim G\over 12 \pi}\int_0^\infty
dp {d\,
\Xi(p)\over dp}.}
The value of this coefficient is different in the three regimes
of ultraviolet regularization.
In the regime dominated by scalar terms $\Xi(\infty)-\Xi(0)=-\pi/2$
and the coefficient takes values $\kappa=\dim G/24$ which are
in agreement with Witten's conjecture (see also \ref\sij
{J.J. van der Bij, R. D. Pisarski, S. Rao,
 Phys. Rev.  D 32 (1985) 2081}\ref\kog{I.A. Kogan,
Phys. Lett. B 256 (1991)
369}.

In the axial regime if $\lambda'>0$, $\Xi$ is a positive function
and $\Xi(\infty)-\Xi(0)=0$. Therefore, there  is not   gravitational
\cs\ term in the induced action. The result is easy to
understand in the
case where the regularized action is purely axial
(i.e. $\lambda=0$), because
 the radiative corrections to the gravitational \cs\ term
vanish in such a case
for all odd orders of perturbation theory. This leads to
conjecture that the
same result holds for $\lambda\neq 0$ but $m\leq 2n+1/2$.
In fact the most
natural result in this regime would be $\kappa=0$ for all
orders of perturbation
theory. However, Axelrod-Singer have calculated the
second order
correction in this regime and found a non-vanishing
contribution \axel. This is a very intriguing result.
A two-loop calculation
in geometric regularization scheme would be very
interesting to verify
whether the above conjecture holds
beyond one loop approximation.
In any case since all odd powers of $\kappa$ in the
expansion \pert\
vanish Witten's conjecture can never be satisfied in
this regime. If $\lambda'<0$ then $\Xi(0)=0$ and also
$\Xi(\infty)=0$, but there exist a singular value of p
where $\Xi$ jumps from $\infty$ to $-\infty$. In such a case
$$\int_0^\infty
dp {d\,
\Xi(p)\over dp}=\pi,$$
and the coefficient $\kappa$ gets a non-null value $$\kappa=
{\dim G\over 12 \pi}.$$
It would be very interesting to calculate the  higher order
corrections to this coefficient and  see if the generalization of
Witten's conjecture also holds for this new universality class.

 In the borderline regime
$m=2n+1/2$ the coefficient $\kappa$ depends  on $\lambda$ and $\lambda'$
\eqn\brek{\kappa={\dim G\over 12 \pi} \arctan{\lambda\over\lambda'}}
in a similar way to the shift of $k$.

Therefore, if $\kappa$ is still related to the central charge
of a two
dimensional conformal theory it has to be a non-rational one.

In this case the non-local term
which counterbalances
the framing dependence  of  gravitational Chern-Simons term
 $G_{\rm cs}$ of the effective action can  not be simply
associated to the standard regularization of the spectral
asymmetry function of the operator $\ast d+ d\ast$.
 The simplest
interpretation of this fact is that    the gravitational
induced effective action can not be   associated in this case to the
gravitational part of standard Atiyah-Patodi-Singer
$\eta$-function \ref\aps{M.F. Atiyah, V.K.Patodi, I.M. Singer,
 Math. Proc. Cambridge Philos. Soc. {77} (1975) 43;
{78} (1975) 405; {79} (1976) 71}. There is  an additional
multiplicative fractional
factor which is regularization  dependent. Whether the same factor
appears in higher order corrections is an interesting conjecture
which requires to be tested beyond the weak  metric
approximation and deserves further investigation.

On the other hand, the dependence on the gauge field of the
effective action or
spectral asymmetry in the above regularizations,  gets also
multiplied by the same factor,  except in the borderline regime,
which makes very interesting to study the spectral asymmetry in this
regime.

\newsec{Discussion and Conclusions}

The  results  analyzed in previous sections show that the three
regimes
of ultraviolet regularization of  the
Chern-Simons action correspond in fact to three different physical
behaviors. In
particular, if the gravitational \cs\ term is eliminated by the
introduction of
a local counterterm in order to get a   metric
independent effective action  a different   framing anomaly is
induced in each regime.
In the axial
regime there is no framing anomaly if $\lambda'>0$ or is a
multiple of ${\dim G/ 12 \pi}$ if $\lambda'<0$. In the
Yang-Mills regime the coefficient of this anomaly is universal and
does not depend, for instance, on the  parameters of the
regulators. This behavior was first pointed out by Witten
\ref\wittten{E. Witten, in {\it Physics and Mathematics of
Strings},   L. Brink ed., World Sci., Singapore (1990)}. In the
borderline regime such a coefficient depends on the weights
$\lambda$ and $\lambda'$
of the axial and scalar regulators and does not correspond
to any previously expected
behavior. This property indicates a possible
connection with
non-rational conformal field theories.

The same analysis holds for the gauge field
effective action as shown in section 4. The effective coupling
constant behaves in a different way for each regime.
It is remarkable the correspondence between the shift of $k$ and the
value of $\kappa$.
Such a relationship is more evident in the borderline case where
both coefficients  depend on the parameters $\lambda$ and
$\lambda'$. However, such a connection does not hold
if we had  chosen
the scalar laplacian $\Delta_A^0$ instead of the Hodge Laplace-Beltrami
operator
\eqn\hodg{\Delta_A=d^\ast_Ad^{\ }_A+d^{\ }_Ad^\ast_A=\Delta_A^0
+\hat{R}(g)+ [F(A),\, .\, ].}
for the regularization of the gluonic action. The difference
between both operators being the  Ricci operator $\hat{R}(g)$ and
the curvature operator $[F(A),\, .\, ]$. In such a case the results for
the first two regimes are unchanged but in the borderline case the
shift of $k$ becomes
\eqn\nuv{k_R= k + {2h^{\vee}\over\pi}\left\lbrack
\arctan{\lambda\over\lambda'}+{\lambda\lambda'\over3(\lambda^2+
\lambda'^2)}
\right\rbrack,}
whereas the coefficient $\kappa$ of the induced Chern-Simons
gravitational term reads
\eqn\breks{\kappa={\dim G\over 12 \pi} \left\lbrack
\arctan{\lambda\over\lambda'}+{16 n \lambda\lambda'\over 5(\lambda^2+
\lambda'^2)}
\right\rbrack.}
The new terms ${\lambda\lambda'/(\lambda^2+
\lambda'^2)}$ come from the boundary
values of integration by parts in the diagrams of the
type (1) in figures 1  and 4 and give  different weights for the
radiative corrections to $k$ and $\kappa$. This fact stresses
the very deep nature of the connection between both quantities in
the natural regularizations formulated purely in terms of
Hodge-Laplace-Beltrami
operators $\Delta_A$. It is only for those
regularizations where the
ratio between both corrections agrees with Witten's conjecture.
It would be very interesting
to investigate if this property also holds beyond one loop approximation.

On the other hand, there is an additional non-analytic contribution to the
effective gauge action in the borderline regime.
It  appears in one loop approximation to the effective action
to compensate the anomalous transformation law of  Chern-Simons
terms under large gauge transformations. The contribution is
similar   to the one which appears in the spectral asymmetry
$\eta$--function of the operator
 $\ast d^{\ }_A+d_A\ast$  induced by the changes of signs in the
spectral flow \ref\red{A.V. Redlich, Phys. Rev. Lett. 52 (1981) 18;
 Phys. Rev. D29 (1984) 2366}
\ref\nuevo{M. Asorey,
F. Falceto,  J.L. L\'opez and G. Luz\'{o}n, In preparation}.
The
coefficient of  such a term also depends on the ultraviolet behavior
of the regularized action, which suggests the   existence
of a
prefactor in the   dependence of the effective action on the
standard $\eta$-function and the
spectral flow. This means that those regularizations which provide
different results  correspond  in fact to different regularizations
of  the spectral asymmetry \nuevo. The
existence  of higher order corrections to this coefficient remains
unclear and  requires further study. A more detailed account of this
phenomenon will be  carried out elsewhere \nuevo.

Although the value  of the effective
coupling
constant can always be modified by  a different choice of
renormalization
scheme,  we remark that the borderline quantization can not be
reduced to the
other two regimes by renormalization. The behavior
of Chern-Simons term under large gauge transformations
implies that the functional integral \siete\ is ill defined unless
the bare coupling constant $k$ is an integer number.
Such a constraint is based on a non-perturbative
effect, because large gauge transformations
map small fields into large gauge fields and, therefore,
they are genuine non-perturbative symmetries.
In consequence, although in perturbation
theory any  local BRST
invariant counterterm is valid, only  counterterms which
preserve the non-perturbative consistency condition
can be added to the bare action. This condition
imposes a very restrictive
constraint on  counterterms which have to preserve the
integer valued character of the bare coupling constant $k$.
In particular, if the
effective value of $k_R$ is not an integer, we cannot
 reduce the
physical behavior of the system to the standard integer
valued case
by a consistent renormalization. Therefore, whereas
 theories defined by
 axial or scalar regularizations might be physically equivalent,
because it  suffices to
consider different values
for the bare coupling $k$ for each case,
the borderline regime yields a new different theory.


Axial and scalar regularizations define quantum theories
in the covariant formalism which agree with the ones obtained by canonical
quantization \ita\LAG.
They have  a finite (although different for the same bare coupling constant
$k$) number of
physical states in a two-dimensional space with the topology of a torus; and
those states are in one-to-one correspondence with
the primary fields of the
corresponding conformal field theories \witten.
If the borderline regularization really defines a new  type of theory,
it must have a different number of states. In this way
it will not be related to rational conformal field theories. Therefore in order
to
elucidate this possibility it is very interesting to
calculate the number of states of the theory
on a two dimensional torus. It can be
obtained in the covariant formalism
 by the analysis
 of the expectation value of an unknotted Wilson
loop $\vev{W^J_U(A)}$ in the spin $J$ representation of $SU(2)$. Following the
methods and  renormalization prescriptions used in Refs.
\ref\lab{E. Guadagnini, N. Martellini, M. Mintchev,  Phys. Lett.{ B 228} (1989)
489: Nucl. Phys.
{ B 330} (1990) 575
\semi M. Alvarez, J.M.F. Labastida, Nucl. Phys. { B 395} (1993) 198}
it is easy to show that with our regularization we get
\eqn\wil{\vev{W^J_U(A)}=2-{\pi^2\over  k^2}+ \xi{2\pi ^2\over
k^3}+\CO\left({1\over k^4}\right),}
where $\xi=k_R-k$ denotes the shift of the effective coupling
constant $k_R$. The absence of frame dependent terms is due to
the fact that in the regularized theory there are not divergences in the
radiative corrections to the Wilson loop. Logarithmic divergences, however,
reappear in  the limit
$\Lambda\rightarrow \infty$. They can be removed by  appropriate counterterms
which in our case have been chosen in such a way that  the whole contribution
of perturbative diagrams with
an isolated propagator vanishes. Such a prescription also removes the ({\it
writhe}) metric dependent part of the Wilson loop. The different terms of \wil\
agree with the
first terms of the perturbative expansion of the formula
\eqn\wii{\vev{W^J_U(A)}={\sin [(2J+1)\pi/(k+\xi)]\over \sin[\pi/(k+\xi)]},}
which generalizes Witten's conjecture \witten.

In  axial ($\xi=0$) and scalar regimes ($\xi=2$) the vacuum
expectation value of the unknotted Wilson loop
$\vev{W^J_U(A)}$ is periodic function in $J$, which indicates that the number
of genuine primary fields of the
corresponding
$\widehat{SU(2)}$ affine algebra  is finite, $k-1$ and $k+1$,
respectively. Moreover, there is in each case a {\it null representation},
$J=\ha (k\mp 1)$  where the value of the
Wilson loop vanishes. It corresponds to a null state in the  module of primary
fields of the corresponding rational conformal field theory.
However,    in the borderline
regime  \wii\ is no longer periodic
in  $J$, because $\xi$ is not an integer\shift, and
 there are not null states which
could be associated to null vectors in  integrable representations of the
$\widehat{SU(2)}$ affine algebra.
Since it is not possible to mod out by  null vectors, the number of
Chern-Simons states when the physical space
is a torus
is infinite, which is a way of interpolating between the dimensions of the
Hilbert spaces of
the theory in the axial ($k-1$) and scalar regimes ($k+1$).

On the other hand the underlying  $SU(2)_q$ quantum
group symmetry, gets a deformation parameter $q= \exp[{2\pi i/(k+\xi)}]$ which
in that regime
is not a root of unity.
Such a behavior points out again that such a theory cannot be related to
rational conformal field theories.

The fact that different regularizations of the theory give
rise to
different quantum theories is quite surprising. Usually,
the effect of the irrelevant terms introduced by the different regularizations
can be absorbed into the renormalization of
coupling constants. However, in the present case this is not possible because
of global constraints. This feature can be associated to the topological
character of the theory. In topological theories,   expectation values  of
topological observables are usually
quantized, therefore,  there exists the possibility of having  irrelevant
perturbations  which    break diffeomorphism
invariance and modify the expectation values in such a way
that the original values
 cannot be recovered by
means of a consistent
renormalization of coupling constants. For instance, in topological quantum
mechanics
on a Riemann surface $\Sigma$ of genus $h$ in the
presence of a magnetic field $A$ with
magnetic charge $k$, the dimension of the space of quantum
states is given
 by the
Riemann-Roch theorem: \
$
\dim {{\cal H}}_k^0= 1-h + k,$
 for  $k> h-1$.
 The quantum hamiltonian is trivial ($H=0$) as
corresponds to a topological theory. However, if we regularize
the theory by means of a metric dependent kinetic term,
$$L(x,\dot x)= {1\over 2 \Lambda} g_{ij}\dot{x}^i\dot{x}^j+
 A_i \dot{x}^i,$$
the Hamiltonian becomes
$H_{\Lambda}={\sevenrm {}\Lambda\over 2}\Delta_A^g$, and the topological limit
$\Lambda\rightarrow \infty$ is governed by
the ground states of $H_{\Lambda}$.
The
quantum Hilbert space of the topological field theory
obtained by this method can have a dimension lower than $1-h+k$, depending on
the symmetries of the background metric
$g$
 of $\Sigma$
\ref\kar{M. Asorey, Geom. Phys. {  11} (1993)
63.}. In particular, this is the case when the metric $g$  breaks the
degeneracy of the ground state of the covariant Laplacian
$\Delta_A^g$. This simple
example shows  how in topological theories
it is possible to obtain different physical theories by means
of different choices of regularization. In some sense, only regularizations
which preserve  certain symmetries
of the theory belong to
 the
 same universality class. For instance, in the previous
 example  only the metrics which are compatible with the magnetic
field $B=dA$, in the sense that they define a K\"ahler structure on $\Sigma$,
lead to the same quantum system.

	In Chern-Simons theory the only regime which
can be characterized by some extra symmetries of the
ultraviolet
regularization is the axial regime when $\lambda=0$.
Only in such a case, the  odd character of the bare action under parity
transformations is  preserved. The effect of this remnant symmetry is to
preserve   the gauge fields   unrenormalized, and consequently the
supersymmetry
associated to Chern-Simons in Landau gauge is recovered in the limit
$\Lambda\rightarrow \infty$ \ref\pigg{F. Delduc, F. Gieres and S.P. Sorella,
Phys. Lett. B253 (1985) 269}.
Such a regime is defined by a special universality
class of regularizations, which give rise to the standard
  Chern-Simons theory defined by canonical quantization after
a finite renormalization of the coupling constant. The
other  regimes introduce
 regularized actions with an hybrid behavior under parity symmetry which is
reflected in the renormalization gauge fields and the
breaking of the special supersymmetry associated to Landau gauge \pigg.
Nevertheless, once  such a renormalization of the gauge field wave function  is
taken into account,
the theory defined by the scalar regime agrees with   the standard theory
defined by canonical quantization.

However,  the borderline regime provides a non-standard quantization
prescription for
Chern-Simons theory in the covariant formalism which is
presumably related to non-rational conformal
field theories. The analysis of such a connection in the canonical
formalism is a challenging open problem.

\bigbreak\bigskip\bigskip\centerline {{\bf Acknowledgements}}
\nobreak We thank A. Morozov and A.A. Slavnov for helpful
discussions.  J. L. L. was supported by a MEC fellowship  (FPI
program)  and G.L. by a CONAI (DGA) fellowship. We also acknowledge
to   CICyT
for partial financial support under grants AEN90-0029 and AEN93-0219.
\bigskip

\vfill\eject

\appendix{A}{}
The  propagator of gluons is obtained from
the restriction of the inverse of the operator
$$
\Gamma_2=-{k \over 4\pi}\left\lbrace -\lambda{\Delta \over \Lambda^2}
\left(1+{\Delta \over \Lambda^2}\right)^{m}
+i*d\left\lbrack1+\lambda'{\Delta \over \Lambda^2}\left(1+
{\Delta \over \Lambda^2}\right)^{2n} \right\rbrack\right\rbrace
$$
to the subspace of transverse fields $A$  $\lbrace d^*A=0 \rbrace$:
$$
\Pi =(1-d(d^\ast d)^{-1}d^\ast)\Gamma_2^{-1}(1-d(d^\ast d)^{-1}d^\ast)
$$
It is given by
$$
{\Pi}_{\mu\nu}^{ab}(p)=-{4\pi\over k}\delta^{ab}{
-\lambda
(1+{p^2/\Lambda^2})^m (p^2\delta_{\nu\mu}-{p_\nu p_\mu})/
\Lambda+
\epsilon_{\mu\rho\nu}p_\rho \left(1+\lambda'(1+{p^2/
\Lambda^2})^{2n}{p^2/\Lambda^2}\right)\over p^2 \lbrack \lambda^2
(1+{p^2/ \Lambda^2})^{2m}{p^2/\Lambda^2}+(1+\lambda'
(1+{p^2/\Lambda^2})^{2n}{p^2/\Lambda^2})^2\rbrack}
$$
The simplest gluon selfinteraction is given by the
three points vertex function, which in  generic regimes
with integer exponents $n$ and $m$ reads
$$
\eqalign{\Gamma_{\mu\nu\rho}^{abc}&(p,q,r)   =
{{ik}\over 24\pi}f^{abc}\epsilon_{\mu\nu\rho} -
{{ik\lambda}\over 48\pi\Lambda}f^{abc} \biggl\lbrace
\epsilon_{\alpha\beta\mu}\epsilon_{\alpha\nu\rho}p_\beta
\left(1+{p^2\over \Lambda^2}\right)^m
 \cr & + {1\over \Lambda^2}
\epsilon_{\sigma\beta\mu}\epsilon_{\alpha\gamma\rho}
p_\beta r_\gamma
(2 r_\nu  \delta_{\sigma\alpha}+q_\sigma \delta_{\alpha \nu} -
q_\alpha \delta_{\sigma \nu}) \rbrack
\sum_{j=0}^{m-1}\left(1+{p^2\over \Lambda^2}\right)^j\left
(1+{r^2\over \Lambda^2}\right)^{m-j-1}  \biggr\rbrace\cr & -
{{ik\lambda'}\over 48\pi\Lambda^2}f^{abc} \biggl\lbrace
\epsilon_{\mu\rho\nu} p^2(1+{p^2\over \Lambda^2})^{2n}  +
\epsilon_{\alpha\beta\mu}\epsilon_{\alpha\nu\gamma}
\epsilon_{\gamma\sigma\rho}p_\beta r_\sigma
\left(1+{p^2\over \Lambda^2}\right)^{n}\left(1+{r^2\over \Lambda^2}
\right)^{n} \cr &+
{1 \over \Lambda^2}
p_\alpha\left[(q_\gamma \epsilon_{\gamma\alpha\mu}\delta_
{\sigma\nu} - q_\sigma\epsilon_{\nu\alpha\mu}
)(r_\sigma r_\rho - r^2 \delta_
{\sigma \rho}) \right.\cr &
+ \left. 2r_\nu r^2\epsilon_{\mu\alpha\rho} \right]\sum_{j=0}^{n-1}
\left(1+{p^2\over \Lambda^2}\right)^{j}\left(1+{r^2\over \Lambda^2}
\right)^{2n-j-1} \cr & -
{1 \over \Lambda^2}\lbrack 2\epsilon_{\mu\alpha\rho}
r_\alpha r_\nu p^2 +
r_\alpha \epsilon_{\gamma\alpha\rho}(q_\gamma\delta_
{\sigma\nu} - q_\sigma\delta_{\gamma\nu})(p_\sigma p_\mu - p^2 \delta_
{\sigma \mu}) \rbrack\cr & \sum_{j=0}^{n-1}
\left(1+{p^2\over \Lambda^2}\right)^{2n-j-1}\left(1+{r^2\over \Lambda^2}
\right)^{j}
\biggr\rbrace +
{\rm{perms}} [ (p,\nu,a),(q,\mu,b),(r,\rho,c)]\cr}
$$
The four point interaction has a similar expression although
considerably
much longer. In order to keep the discussion in  reasonable terms
we omit the explicit expression which on the other hand can be
derived by standard methods.

There is a subtle point in
the analysis of Feynman rules for
the intermediate case. There, since $m=2n+1/2$
the regularization involve pseudodifferential
operators in the regulators of gauge fields. If we assume
$n$ to be an integer then the pseudodifferential operators
appear in the Yang-Mills like part of the
regularized action
$$( F(A),
(I + {\Delta_A/ \Lambda^{^2}})^{m}  F(A))
$$
We assume the standard definition of pseudodifferential
operators and
we  proceed by expanding the corresponding perturbative
expression in a momentum basis. If we denote by
$\CF(A)$ the pseudodifferential operator
$(I + {\Delta_A/ \Lambda^{^2}})^{m}$
the corresponding Feynman rules can be derived from the
following perturbative expansion:
$${\eqalign{\CF(&A)^{ac}_{\mu\rho}(p,-p-q)=
\delta^{ac}\delta_{\mu\rho}\delta(q)\CF_0(p)+ {1\over \Lambda^2}\biggl\{
 i f^{abc} A^b_{\nu}(q)\lbrack
(2p+q)_\nu\delta_{\mu\rho}-q_\mu\delta_{\nu\rho}+
q_\rho\delta_{\nu\mu}\rbrack
\cr
&+
\int
{d^3 q'\over (2\pi)^3}A^b_{\nu}(q-q')A^{b'}_{\nu'}
(q')(f^{bb'e}f^{aec}\delta_{\mu\nu'}
\delta_{\rho\nu}-
f^{abe}f^{eb'c}\delta_{\mu\rho}\delta_{\nu\nu'})
\biggr\}
{\CF_0(p+q)-\CF_0(p)\over (p+q)^2-p^2} \cr & -
{1\over \Lambda^4}f^{abe}f^{eb'c}
\int
{d^3 q'\over (2\pi)^3}A^b_{\nu}(q-q')A^{b'}_{\nu'}
(q')
\lbrack
(2p+q-q')_\nu\delta_{\mu\sigma} -(q_\mu-q'_\mu)\delta_{\nu\sigma}\cr  & +
(q_\sigma-q'_\sigma)\delta_{\nu\mu}\rbrack
\lbrack
(2p+2q-q')_{\nu'}\delta_{\sigma\rho}-{q'}_\sigma
\delta_{\nu'\rho}+{q'}_\rho
\delta_{\nu'\sigma}\rbrack
\, {1\over p^2-(p+q)^2 }\cr
&
\phantom{+-}\biggl\lbrack
{\CF_0(p+q-q')-\CF_0(p)\over (p+q-q')^2-p^2}-
{\CF_0(p+q-q')-\CF_0(p+q)\over (p+q-q')^2-
(p+q)^2}\biggr\rbrack +\CO{(A^3)}
,}}
$$
where $\CF_0(p)=(1+p^2/\Lambda)^m$.

Feynman rules for the interaction of nuclear and metric
ghosts can be found
in \gr\ (see also \ref\tesis{F. Falceto, {\it Regularizaci\'on
Geom\'etrica de Teor\'{\i}as Gauge}, Ph. D. Dissertation,
 Zaragoza University (1989)}).

\appendix {B}{}
Using the Feynman rules of Appendix A it is possible to calculate
the radiative corrections to the two-point function generated by
gluonic loops (diagrams (1) and  (2) of Fig. 1). The scalar
parts of those contributions are
\eqn\gluons{\eqalign {^{\phantom{}(1)}_{\phantom{(}
s}
\Gamma_{\mu \nu}^{ab}(q)  = &
{2h^{\vee}\over 3\pi^2}\, (m+1)^2\,\Omega\,\delta^{ab}\,
\delta_{\mu\nu}
\cr & +
{ h^{\vee}\over 3\pi^2}\,\Lambda\, I(n,m)\,\delta^{ab}\,
\delta_{\mu\nu}
  \cr & -
{h^{\vee}\over 8\pi^2}\, (m+1)^2 \, \delta^{ab}\vert q \vert\left
(
\delta_{\mu\nu} +
{q_{\mu}q_{\nu} \over q^2} \right )  \cr & +
\CO(\Omega^{-1},\Lambda^{-1}), \cr}
}
and
\eqn\gluonss{\eqalign{^{\phantom{}(2)}_{\phantom{(}
s}
\Gamma_{\mu \nu}^{ab}(q) = & -
{ h^{\vee}\over 3\pi^2}(2m^2+5m+2)\,\Omega\,\delta^{ab}\,
\delta_{\mu\nu}
\cr & - {h^{\vee}\over 3\pi^2}\,\Lambda\, I(n,m)\,
\delta^{ab}\,\delta_{\mu\nu}
\cr &+ \CO(\Omega^{-1},\Lambda^{-1}), \cr }
}
if $m\geq 2n+1/2$
\eqn\gluons{\eqalign { ^{\phantom{}(1)}_{\phantom{(}
s}
\Gamma_{\mu \nu}^{ab}(q)
 = &
{2h^{\vee}\over 3\pi^2}\, (2n+3/2)^2\,\Omega\,\delta^{ab}\,
\delta_{\mu\nu}
\cr & +
{ h^{\vee}\over 3\pi^2}\,\Lambda\, I(n,m)\,\delta^{ab}\,
\delta_{\mu\nu}  \cr & -
{h^{\vee}\over 32\pi^2}\, (4n+3)^2 \, \delta^{ab}\vert q
\vert\left (
\delta_{\mu\nu} +
{q_{\mu}q_{\nu} \over q^2} \right )  \cr & +
\CO(\Omega^{-1},\Lambda^{-1}), \cr}
}
and
\eqn\gluonss{\eqalign{^{\phantom{}(2)}_{\phantom{(}
s}
\Gamma_{\mu \nu}^{ab}(q)
 = & -
{ h^{\vee}\over 3\pi^2}(8n^2+14n+5)\,\Omega\,\delta^{ab}\,
\delta_{\mu\nu}
\cr & - {h^{\vee}\over 3\pi^2}\,\Lambda\, I(n,m)\,
\delta^{ab}\,\delta_{\mu\nu}
\cr &+ \CO(\Omega^{-1},\Lambda^{-1}), \cr }
}
when $m\leq 2n+1/2$,
with
$$
I(n,m)=\int_{0}^\infty
{\tau(p)\over \rho(p)}dp
$$
\eqn\sthree{\rho(p)=\lambda^2 p^2(1+p^2)^{2m}+
\lbrack 1+\lambda' p^2(1+p^2)^{2n}\rbrack^2,
}
and
$$
\eqalignno {\tau(p)= & 2\lambda'\lbrack 1+\lambda'
p^2(1+p^2)^{2n}\rbrack
\lbrack 5+2p^2(5+9n) \rbrack p^2 (1+p^2)^{2n-2}- \cr & -
\lambda^2 \left\lbrack mp^2(1+p^2)^{2m-2}\lbrack (2m+5)+
(4m+3)p^2 \rbrack +
(2m^2 + 5m + 2) \right\rbrack. \cr}
$$
The pseudoscalar contributions to $\Gamma_{\mu\nu}^{a b}$
generated by gluonic loops (diagrams (1)--(2) of fig. 1) are given
by
\eqn\stwo{^{\phantom{}(1)}_{\phantom{(}
\ast}
\Gamma_{\mu \nu}^{ab}(q)= {
h^{\vee}\over3\pi^2}\int_{0}^\infty {A(p)\over
\rho(p)^2}dp\ \delta^{a
b}\epsilon_{\mu\sigma\nu} q^\sigma, }
and
\eqn\sone{ ^{\phantom{}(2)}_{\phantom{(}
\ast}
\Gamma_{\mu \nu}^{ab}(q)=
{ h^{\vee}\over3\pi^2}\int_{0}^\infty {\bar{A}(p)\over \rho(p)}dp\
\delta^{a
b}\epsilon_{\mu\sigma\nu} q^\sigma, }
respectively, where
\eqn\sfive{A(p)=B(p)\rho(p)+ C(p){d\rho(p) \over
dp}+ D(p){d\phi(p) \over dp},
 }
\eqn\sfour{\eqalign {\bar{A}(p)=& 2\lambda R (p) +
3 S (p) R (p) - 5 S (p)  - m\lambda\lbrack 7p^2+(5+2m)p^4\rbrack
  (1+p^2)^{m-2} R(p) \cr & +
  S (p)
2n\lambda' p^4 \lbrack 9+(7+4n)p^2\rbrack(1+p^2)^{2n-2},
 \cr }}
and
$$\eqalign{B(p)
= &-\lambda R (p) - S(p) R(p)+
3 S(p) + {d (S(p)p)\over dp}
+3 p\lambda {dR(p)\over dp}\cr
&+(6m-8n)\lambda' p^4 (1+p^2)^{2n-1} S(p) \cr} $$
$$\eqalign{C(p)=&-2p\lambda R(p)+
{1 \over 2}p{d(S(p)p)\over dp}+pS (p) \bigg\lbrack {3\over2}+
m\lambda' p^4 (1+p^2)^{2n-1} -{1 \over 2}p{dR(p)\over dp}
\bigg\rbrack \cr}$$
$$D(p)= \lambda
p^2 S(p) R^2(p) \lbrack 2+(1+p^2)^m \rbrack,
$$
$$
S(p) = \lambda (1+p^2)^m
\qquad
R(p) = 1+\lambda' p^2(1+p^2)^{2n}\qquad
\phi(p)=p{S(p) \over R(p)}.
$$
In  the first two terms of \sfive\ into \stwo\
one $\rho(p)$--factor can be removed  from the denominator by
 integration by parts.  Adding
the result  to the contribution \sone\ of the diagram
(2) we obtain
\eqn\sseven{\bigg\lbrack\int_{0}^\infty {\theta(p)\over
\rho(p)}dp+\chi_1\bigg\rbrack \delta^{a
b}\epsilon_{\mu\sigma\nu} q^\sigma{
h^{\vee}\over3\pi^2}, }
with
$$\theta(p)  =-\lambda R(p)+\lambda p {dR(p)
\over dp}+2 S(p)+ R(p) {d(S(p)p)\over dp}-
 pS(p){dR(p)\over dp}
$$
and
$$
\eqalign{
\chi_1=\lbrack\lim_{p\to\infty}-\lim_{p\to 0}\rbrack
{1 \over \rho(p)}& \bigg\lbrack{1\over 2}p^2 S(p){d(R(p))\over dp}+
{1\over 2}p S(p)R(p)-2pS(p)\cr
&-{1\over 2}p R(p){d(S(p)p)\over dp}
+2\lambda p R(p)\bigg\rbrack\cr}
$$
%
  The remaining term of \sfive\ after integration by parts yields
\eqn\seight{
{ h^{\vee}\over 6\pi^2}\bigg\lbrack\int_{0}^\infty  {\xi(p)
\over  \rho(p) }\, dp-\chi_2\bigg\rbrack \delta^{a
b}\epsilon_{\mu\sigma\nu} q^\sigma,}
where
$$
\eqalign {\xi(p) & =2\lambda R(p)-2\lambda p
{dR(p)\over dp}
+ R(p){d(S(p)p)\over dp} - p S(p) {dR(p)\over dp}
\cr }
$$
and
\eqn\schi{
\chi_2=\lbrack\lim_{p\to\infty}-\lim_{p\to 0}\rbrack
{\lambda p\lbrack 2+(1+p^2)^m \rbrack \lbrack
1+\lambda'p^2(1+p^2)^{2n} \rbrack \over
\rho(p)} .}
The
global contribution of gluonic loops to the two point function
is obtained by summing up the contributions \sseven\ and \seight,
\eqn\snine{ ^{}\Pi_{\mu\nu}^{a b}(q)={h^{\vee}\over3\pi^2}
 \int_{0}^\infty dp\, {\Sigma(p) \over
\rho(p)}\delta^{a
b}\epsilon_{\mu\sigma\nu} q^\sigma}
where
\eqn\sten{\eqalign{
\Sigma(p)=&2 S(p)+{3 \over2} \bigg\lbrack R(p)
{d(S(p)p)\over dp}-pS(p) {dR(p)\over dp} \bigg\rbrack
.\cr}}
Notice that the calculation in the borderline case which involves
pseudodifferential operators reduces to the analytic continuation
in the regulating parameters $n$ and $m$ of the other two cases.

Finally, the radiative corrections to the vacuum polarization tensor
generated by loops of metric and nuclear ghosts loops are
purely scalar because their interactions  do not involve
pseudoscalar couplings. They
read
\eqn\ghost{\eqalign { \Pi'{}_{\mu \nu}^{ab}(q)  = &
+{h^{\vee}\over 3\pi^2}\, (m_2-1)\,\Omega\,\delta^{ab}\,
\delta_{\mu\nu}
\cr &
 +
{h^{\vee}\over 8\pi^2}\, [(m_0)^2 +(m_2-m_1)^2+(m_1-m_0)^2]\,
\delta^{ab}\vert q \vert\left  (
\delta_{\mu\nu} +
{q_{\mu}q_{\nu} \over q^2} \right )  \cr & +
\CO(\Omega^{-1},\Lambda^{-1}).\cr}
}
\listrefs

\end

\figures
\fig {1.} { Radiative corrections to the vacuum polarization
involving gluon loops.}
\fig {2.} { Radiative corrections to the vacuum polarization
involving ghosts loops.}
\fig {3.} {  One loop contributions to the   self-energy of the
Faddeev-Popov ghosts (diagram (1)), and
  to the
3-vertex gluon-ghost interaction (diagrams (2) and (3)).}
\fig {6.} { One loop   contributions to the graviton 2-point function
involving gluon loops.}
\fig {4.} { The three regimes of ultraviolet behavior of the
$\phi$ function for $\lambda'>0$. $p_0$ is the critical
point where
the function $\phi$ reachs its maximal value in the regime
$m<2n+{1\over 2}$.}
\fig {5.} { The three regimes of ultraviolet behavior of
the $\phi$ function for $\lambda'<0$. The infinite gap at $p_\infty$
corresponds to a zero value of the pseudoscalar leading
term which appears in all  regimes. }

\end
The four point interaction is given by the following expression
$$\eqalign{ -{\lambda' k\over24\pi\Lambda^2}
f^{abe}&f^{ecd}\biggl\lbrack
{1\over 8\Lambda^2}\epsilon_{\alpha\beta\mu}
\epsilon_{\alpha\gamma\delta}\epsilon_{\delta\eta\sigma}
p_\beta p_\gamma s_\eta\biggl\lbrack -\delta_{\nu \rho}
\sum_{j=0}^{n-1}
\left(1+{p^2\over\Lambda^2}\right)^{n+j} \left
(1+{s^2\over\Lambda^2}\right)^{n-j-1} \cr &
+{4\over\Lambda^2} s_\rho  (r+s)_\nu
\sum_{j=0}^{n-2}\sum_{l=0}^{n-j-2}\left(1+{p^2\over\Lambda^2}
\right)^{n+j}
\left(1+{(r+s)^2\over\Lambda^2}\right)^l\left
(1+{s^2\over\Lambda^2}\right)^{n-j-l-2}\biggr\rbrack \cr & +
{1\over 8\Lambda^2}\epsilon_{\alpha\beta\mu}
\epsilon_{\alpha\gamma\delta}\epsilon_{\delta\eta\sigma}
p_\beta s_\gamma s_\eta\biggl\lbrack -\delta_{\nu \rho}
\sum_{j=0}^{n-1}
\left(1+{p^2\over\Lambda^2}\right)^j \left
(1+{s^2\over\Lambda^2}\right)^{2n-j-1} \cr & +
{4\over\Lambda^2} s_\rho  (r+s)_\nu
\sum_{j=0}^{n-2}\sum_{l=0}^{n-j-2}\left(1+
{p^2\over\Lambda^2}\right)^j
\left(1+{(r+s)^2\over\Lambda^2}\right)^l
\left(1+{s^2\over\Lambda^2}
\right)^{2n-j-l-2} \biggr\rbrack\biggr\rbrace
 \cr &  +
{\rm{perm}} [ (p,\mu,a),(q,\nu,b),(r,\rho,c),
(s,\sigma,d] .\cr }
$$
$$
\eqalign{\Gamma
_{\mu\nu\rho\sigma}^{abcd} & (p,q,r,s)=-{\lambda
k\over24\pi\Lambda}
f^{abe}f^{ecd}\biggl\lbrace
{1\over32}(\delta_{\mu\rho}\delta_{\nu\sigma} -\delta_{\mu\sigma}
\delta_{\nu\rho})\left(1+{(p+q)^2\over \Lambda^2}\right)^m \cr & +
{1\over8\Lambda^2}(p_\rho\delta_{\sigma\mu}-p_\sigma\delta_{\rho\mu})
(r+s)_\nu\sum_{j=0}^{m-1}\left(1+{p^2\over\Lambda^2}\right)^j
\left(1+{(r+s)^2\over\Lambda^2}\right)^{(m-j-1)} \cr & +
{1\over8\Lambda^2}(s_\mu\delta_{\sigma\nu}-s_\nu\delta_{\sigma\mu})
s_\rho\sum_{j=0}^{m-1}\left(1+{s^2\over\Lambda^2}\right)^j
\left(1+{(p+q)^2\over\Lambda^2}\right)^{(m-j-1)} \cr & +
{1\over8\Lambda^2}(p_\alpha s_\alpha\delta_{\sigma\mu}-p_\sigma
s_\mu) \biggl\lbrack
\delta_{\nu\rho}\sum_{j=0}^{m-1}\left(1+{p^2\over\Lambda^2}\right)^j
\left(1+{s^2\over\Lambda^2}\right)^{(m-j-1)}  \cr & + (r+s)_\nu
s_\rho{4\over\Lambda^2}\sum_{j=0}^{m-2}\sum_{l=0}^{m-j-2}
\left(1+{p^2\over\Lambda^2}\right)^j\left(1+{(r+s)^2\over\Lambda^2}
\right)^l
\left(1+{r^2\over\Lambda^2}\right)^{(m-j-l-2)}\biggr\rbrack\biggr
\rbrace \cr &
-{\lambda' k\over24\pi\Lambda^2}
f^{abe}f^{ecd}\biggl\lbrace
{1\over32}\lbrack
\epsilon_{\nu\rho\sigma}(r+s)_\mu  \cr  & -\epsilon_{\mu\rho\sigma}
(r+s)_\nu\rbrack \left(1+{(r+s)^2 \over \Lambda^2}\right)^{2n}
\cr & + {1\over
8\Lambda^2}\epsilon_{\alpha\beta\mu}
\epsilon_{\alpha\gamma\delta}\epsilon_{\delta\rho\sigma}
p_\beta p_\gamma
(r+s)_\nu\sum_{j=0}^{n-1}
\left(1+{p^2\over\Lambda^2}\right)^{n+j}\left(1+{(r+s)^2\over
\Lambda^2}
\right)^{n-j-1} \cr & + {1\over 16}\epsilon_{\alpha\beta\mu}
\epsilon_{\alpha\nu\gamma}\epsilon_{\gamma\rho\sigma}
p_\beta \left(1+{p^2\over\Lambda^2}\right)^n\left
(1+{(r+s)^2\over\Lambda^2}\right)^n \cr & +
{1\over 8\Lambda^2}\epsilon_{\alpha\beta\mu}
\epsilon_{\alpha\gamma\delta}\epsilon_{\delta\rho\sigma}
p_\beta (r+s)_\gamma (r+s)_\nu\sum_{j=0}^{n-1}
\left(1+{p^2\over\Lambda^2}\right)^j\left(1+{(r+s)^2\over\Lambda^2}
\right)^{2n-j-1}\cr & +
{1\over 8\Lambda^2}\epsilon_{\alpha\mu\nu}
\epsilon_{\alpha\beta\gamma}\epsilon_{\gamma\delta\sigma}
(r+s)_\beta s_\rho s_\delta \sum_{j=0}^{n-1}
\left(1+{(r+s)^2\over\Lambda^2}\right)^{n+j}
\left(1+{s^2\over\Lambda^2}\right)^{n-j-1} \cr & +
{1\over 16}\epsilon_{\alpha\mu\nu}
\epsilon_{\alpha\rho\beta}\epsilon_{\beta\gamma\sigma}
s_\gamma \left(1+{s^2\over\Lambda^2}\right)^n\left
(1+{(r+s)^2\over\Lambda^2}\right)^n \cr & +
{1\over 8\Lambda^2}\epsilon_{\alpha\mu\nu}
\epsilon_{\alpha\beta\gamma}\epsilon_{\gamma\delta\sigma}
s_\beta s_\rho s_\delta \sum_{j=0}^{n-1}
\left(1+{(r+s)^2\over\Lambda^2}\right)^j\left
(1+{s^2\over\Lambda^2}\right)^{2n-j-1} \cr & +
{1\over 4\Lambda^2}\epsilon_{\alpha\beta\mu}
\epsilon_{\alpha\nu\gamma}\epsilon_{\gamma\delta\sigma}
p_\beta s_\rho s_\delta \left(1+{p^2\over\Lambda^2}\right)^n
\sum_{j=0}^{n-1}
\left(1+{(r+s)^2\over\Lambda^2}\right)^j\left
(1+{s^2\over\Lambda^2}\right)^{n-j-1} \cr & +
{(s+r)_\nu\over 4\Lambda^2}\epsilon_{\alpha\beta\mu}p_\beta\biggl
\lbrack
\epsilon_{\alpha\rho\gamma}\epsilon_{\gamma\delta\sigma}
s_\delta \left(1+{s^2\over\Lambda^2}\right)^n \sum_{j=0}^{n-1}
\left(1+{p^2\over\Lambda^2}\right)^j\left
(1+{(r+s)^2\over\Lambda^2}\right)^{n-j-1} \cr &
+{2\over \Lambda^2}
\epsilon_{\alpha\gamma\delta}\epsilon_{\delta\eta\sigma}
s_\rho s_\eta(r+s)_\gamma  \sum_{j,l=0}^{n-1}
\left(1+{p^2\over\Lambda^2}\right)^j
\left(1+{(r+s)^2\over\Lambda^2}\right)^{n-j+l-1}\left
(1+{s^2\over\Lambda^2}\right)^{n-l-1} \biggr\rbrack \biggr\rbrace
\cr}
$$